\documentclass[preprint,12pt]{elsarticle}

\usepackage{lineno}
\modulolinenumbers[1]

\usepackage{siunitx}
\usepackage{amsmath,amssymb}
\usepackage{bm}
\usepackage{stmaryrd} % double brackets
\usepackage{mathrsfs} % macros mathscr
\usepackage{graphicx}
\usepackage{caption}
\usepackage{subcaption}
\usepackage{enumerate}
\usepackage{tikz}
\usetikzlibrary{backgrounds}
\usepackage{xcolor}
\usepackage[colorlinks=true, linkcolor=blue, citecolor = blue]{hyperref}

\journal{Computers and Fluids}
\begin{document}

\begin{frontmatter}
% \title{Numerical modeling of soft particles in microchannel flows\tnoteref{mytitlenote}}
% \tnotetext[mytitlenote]{Fully documented templates are available in the elsarticle package on \href{http://www.ctan.org/tex-archive/macros/latex/contrib/elsarticle}{CTAN}.}
\title{An isogeometric boundary element method for soft particles flowing in microfluidic channels}

% %% Group authors per affiliation:
% \author{Elsevier\fnref{myfootnote}}
% \address{Radarweg 29, Amsterdam}
% \fntext[myfootnote]{Since 1880.}
% 
% %% or include affiliations in footnotes:
% \author[mymainaddress,mysecondaryaddress]{Elsevier Inc}
% \ead[url]{www.elsevier.com}
% 
% \author[mysecondaryaddress]{Global Customer Service\corref{mycorrespondingauthor}}
% \cortext[mycorrespondingauthor]{Corresponding author}
% \ead{support@elsevier.com}
% 
% \address[mymainaddress]{1600 John F Kennedy Boulevard, Philadelphia}
% \address[mysecondaryaddress]{360 Park Avenue South, New York}
                                                                                                                           
\author[M2P2]{J.~M. Lyu}
\author[M2P2]{Paul G. Chen\corref{mycorrespondingauthor}}
\cortext[mycorrespondingauthor]{Corresponding author}
\ead{gang.chen@univ-amu.fr}
\author[IRPHE]{G. Boedec}
\author[LRP]{M. Leonetti}
\author[M2P2]{M. Jaeger}
%\ead{marc.jaeger@centrale-marseille.fr}                        
\address[M2P2]{Aix Marseille Univ, CNRS, Centrale Marseille, M2P2, Marseille, France}
\address[IRPHE]{Aix Marseille Univ, CNRS, Centrale Marseille, IRPHE, Marseille, France}
\address[LRP]{Univ. Grenoble Alpes, CNRS, Grenoble INP, LRP, Grenoble, France}

\begin{abstract} 
Understanding the flow of deformable particles such as liquid drops, synthetic capsules and vesicles, and biological cells confined in a small channel is essential to a wide range of potential chemical and biomedical engineering applications. Computer simulations of this kind of fluid-structure (membrane) interaction in low-Reynolds-number flows raise significant challenges faced by an intricate interplay between flow stresses, complex particles' interfacial mechanical properties, and fluidic confinement.  Here, we present an isogeometric computational framework by combining the finite-element method (FEM) and boundary-element method (BEM) for an accurate prediction of the deformation and motion of a single soft particle transported in microfluidic channels. The proposed numerical framework is constructed consistently with the isogeometric analysis paradigm; Loop's subdivision elements are used not only for the representation of geometry but also for the membrane mechanics solver (FEM) and the interfacial fluid dynamics solver (BEM). We validate our approach by comparison of the simulation results with highly accurate benchmark solutions to two well-known examples available in the literature, namely a liquid drop with constant surface tension in a circular tube and a capsule with a very thin hyperelastic membrane in a square channel. We show that the numerical method exhibits second-order convergence in both time and space. To further demonstrate the accuracy and long-time numerically stable simulations of the algorithm, we perform hydrodynamic computations of a lipid vesicle with bending stiffness and a red blood cell with a composite membrane in capillaries. The present work offers some possibilities to study the deformation behavior of confining soft particles, especially the particles' shape transition and dynamics and their rheological signature in channel flows.
\end{abstract}


\end{frontmatter}

% \linenumbers
 
\section{Introduction}
Microfluidics refers to devices and methods for the manipulation of fluids and immersed objects  inside channels with dimensions of tens to hundreds of micrometers~\cite{Stone_2004}. Understanding the flow of soft or deformable micro-particles (e.g., viscous droplets, artificial capsules and vesicles, and biological cells, etc.) in confined channels is essential for many multiphase microfluidic applications like control and application of droplets (or bubbles)~\cite{Shui_2007}, and particle/cell sorting~\cite{Geislinger_2014,WyattShieldsIV_2015}. It is also fundamental to biological material transport through the microcirculation such as red blood cells (RBCs)~\cite{SECOMB2013470,Tomaiuolo2009} and drug delivery~\cite{Liu_2017}. It is therefore essential to study how soft particles deform under the action of flow stresses in a small (confined) channel and what effect these stresses can have on the transport of the deformable particles and associated processes. Of particular interest are the particle's translational velocity, the overall hydraulic resistance in a given channel containing suspended particles, and the disturbed flow field induced by the presence of suspended particles~\cite{Baroud_2010,Anna_2016}.

The dynamical behavior of these soft objects in an imposed flow of an external fluid exhibits distinct characteristics. The mechanical properties of the interfacial (membrane) composition play a key role in this dynamics. The simplest deformable object is a clean, surfactant-free drop, which is characterized by its interfacial material property -- namely interfacial tension. Other soft entities, however, have increased structural complexity leading to more complex mechanical properties. For instance, a synthetic, liquid-filled capsule can be thought of as a drop enclosed by a solid polymerized membrane that resists shear and area deformation, while a lipid vesicle is a drop enclosed by an inextensible, fluidic membrane resisting bending. Shearing and stretching deformations of a vesicle are negligibly small compared to bending ones. The surface-area incompressibility of the lipid membrane is ensured dynamically by a Lagrange field -- namely the membrane tension which is not a material property but is flow dependent, analogous to pressure for three-dimensional incompressible flows. Biological cells like RBCs have a more complex architecture. The RBCs membrane consists of a lipid bilayer (vesicle-like) and an underlying membrane-associated cytoskeleton (capsule-like).  As such, vesicles and capsules have often served as a model system to mimic RBCs~\cite{Vlahovska_2013}. These soft entities (i.e., drops, capsules, vesicles, and RBCs) are considered in the present work. To facilitate subsequent description, we shall make no difference between interface and membrane.

The soft particles flowing in microfluidic channels is essentially a fluid-structure interaction (FSI) problem involving highly deformable membranes (interfaces). The small size and the low speed of microfluidics mean that the viscous forces predominate the inertial forces and the linear Stokes equations can, therefore, be applied to describe such low-Reynolds-number flow ($Re$ usually much less than unity)~\cite{Stone_2004}. Despite the linearity of the flow, FSI problems at small scales are highly nonlinear due to the highly nonlinear nature of the deformation of the soft objects and their constitutive models~\cite{Barthes_2016,Freund_2014}. Moreover, a significant difficulty arises from the hydrodynamic interaction between the microfluidic wall and the object's surface, especially at conditions of high confinement where confinement-induced viscous friction plays a dominant role in the forces exerted on the soft object~\cite{Trozzo_JCP_2015,Chen_2019}. It is, therefore, a computationally challenging task to solve this kind of FSI problems with fidelity.

There are a variety of different computational strategies to solve FSI problems. Recently developed approaches can be broadly classified into three categories (see, e.g., the reviews~\cite{Barthes_2016} for capsules,~\cite{Abreu_2014} for vesicles, and \cite{Freund_2014} for RBCs), namely bulk mesh-based methods~\cite{LI2007Augmented,Mendez_2014,YE2015Numerical, ZHANG2019Lattice}, particle-based methods~\cite{Noguchi2005Shape,FedosovCaswell2010b,Lanotte2016red}, and boundary-element methods (BEMs). 
 
The BEM offers very high accuracy compared to other methods in the prediction of the dynamics of deformable particles immersed in inertialess Newtonian flows. BEM's theory and its practical implementation are well-described in~\cite{Pozrikidis_1992}.
Its efficiency has been demonstrated in the simulation of drops~\cite{Lac_JFM2009,Nagel2015Boundary,JoneidI2015Isogeometric,Gounley_JFM_2016}, capsules~\cite{Hu_JFM_2012}, vesicles~\cite{Boedec_JCP2011,Farutin_JCP_2014,Boedec_JCP_2017,barakat_shaqfeh_2018b}, and RBCs~\cite{Zhao_JCP2010}. The BEM has a notable advantage over domain discretization methods as it leads to a reduction in dimensionality;  the flow equations are solved only for the unknown stress and velocity fields at the domain boundaries and at evolving interfaces. This restriction of the discretization only to the boundaries and interfaces greatly improve the computational efficiency when studying wall-deformable object interactions.   A prominent example of such benefits is to resolve the drainage fluid of thin liquid film between the object surface and the channel wall at high fluidic confinement, where the object occupies a large proportion of the channel cross-section. Regardless of how close the object is to the channel wall, and how their distance changes over time, the film thickness needs only to be considered when it comes to determining the size of the mesh elements on the membrane and at the channel wall. In their numerical studies on vesicle dynamics confined in tube flow, Refs.~\cite{Trozzo_JCP_2015, Chen_2019} provided an estimate of a typical element size which is about half of this thickness, in the region of the liquid film.
This very worthy property of the BEM eliminates any problem of volume mesh topology related to the movement of an interface or error induced by the interpolation of physical fields to it.
However, it should be tempered by the fact that the matrix system to be solved in this method is characterized by full and not sparse matrices as in most other methods. In other words, an increase in the degrees of freedom in modeling is seriously more penalizing in terms of computation time.
Therefore, any discretization method that increases the accuracy of the numerical representation of surfaces and domain boundaries, while also optimizing the degrees of freedom of this representation, is of great interest to the BEM.

 In structural design, the most convenient and widespread way to represent a surface is to use a triangular mesh. It enables any surface shape to be represented and allows for the development of adaptation algorithms and local mesh refinement. The simplest triangular elements provide linear interpolation per piece of the surface, as well as all physical fields that use the same approximation (referred to as isoparametric in finite element language). That is the strategy adopted in \cite{Boedec_JCP2011}.  The second strategy makes use of the quadratic triangular elements~\cite{Ramanujan_1998}, which allows for an improved accuracy without increasing the degrees of freedom, but at the cost of a slightly more complex numerical implementation.
In both cases, as with all Lagrange elements, the approximation can only be $C^0$ continuous between elements, thus the spatial derivatives are discontinuous across them. The regularity of representation could be increased to $C^1$ using Hermitian elements. Their implementation in 2D (or axisymmetric), though more complex, remains affordable, the extension to 3D is much more problematic, if not impossible. Moreover, we are not aware of any studies using this option.

Being limited to approximations of $C^0$ is particularly penalizing for other obvious reasons.
A direct computation of the membrane bending force requires a $C^4$ representation of the membrane geometry since the bending force contains the fourth-order derivative of the position vector~\cite{Zhong_can_1989}. That is why some local surface reconstruction techniques have been developed to compute the Laplace-Beltrami operator~\cite{Farutin_JCP_2014,Guckenberger_CPC_2016}.  An interesting alternative approach involves the use of differential geometry techniques, as demonstrated by~\cite{Boedec_JCP2011}. Now, one of the advantages of representation by finite element is that the mesh smoothness requirements can be eased from $C^4$ to $H^2$ if formulating the interfacial mechanical problem in weak form~\cite{Cirak_2000}.
In the mathematical field of functional analysis, the Sobolev space $H^2$ represents square-integrable functions whose first- and second-order derivatives are themselves square-integrable. It means that with weak formulation a $C^1$ finite element approximation is only required to compute the bending force since the second derivatives are then piecewise continuous.

The representation of surfaces is a longstanding issue in computer graphics. Exploiting the highly accurate interpolation functions developed in this field, like splines and NURBS (Non-Uniform Rational Basis Splines), represents a recent breakthrough in finite element analysis.
Not only, it eases the connection with computer-aided design (CAD), but it opens the way to increased regularity of finite element approximation in general. All the physical fields involved in a problem to be solved with finite elements benefit from the same highly accurate representation as to the geometry.
This rapidly developing trends in finite element analysis are thus named isogeometric analysis or IGA~\cite{Hughes2005Isogeometric}.

In the IGA framework, the surface subdivision is well suited when only the domain boundaries and evolving interfaces have to be considered, as in the BEM. Its use for finite element analysis with meshes made of triangular elements has been made possible thanks to the work of Loop~\cite{Loop_1987}, with a first application to shell analysis by Cirak et al.~\cite{Cirak_2000}.
IGA-Loop guarantees almost everywhere a $C^2$ approximation, except at a few irregular nodes where it's only $C^1$.
An IGA-BEM model with Loop subdivision for soft particles in the Stokes flows is proposed in~\cite{Boedec_JCP_2017}. Its efficiency has been demonstrated to simulate free-space FSI problems involving drops, capsules, and vesicles in a unified numerical modeling framework. 
Its superior accuracy as compared to previous methods to compute geometric properties like curvature was confirmed in~\cite{Guckenberger_CPC_2016}. A coupled IGA-BEM and shell approach is seen as a promising way to study the interaction between thin elastic structures and Stokes flows and is attracting growing attention in the soft particle's community~\cite{Maestre_CMAME_2017,barakat_shaqfeh_2018b,Bartezzaghi_CMAME_2019}.

The algorithm developed in~\cite{Boedec_JCP_2017} is limited to free-space Stokes flows. In ~\cite{barakat_shaqfeh_2018b,barakat_shaqfeh_2019} three-dimensional computations of a vesicle flowing in a circular and rectangular duct (without bending force) have been reported.
However, whereas the vesicle surface benefited from the increased accurate representation of the Loop subdivision, the unknown physical fields did not. Instead, they were represented by a piecewise linear interpolation, like earlier models. In that regard, the IGA spirit cannot be advocated.

In this paper, we extend the previous work~\cite{Boedec_JCP_2017} on soft particles in unbounded Stokes flows to confined soft particles transported in microfluidic channels. Both the surface shape of soft objects and the wall boundary of channels are discretized with Loop's subdivision scheme.
Most importantly, Loop elements are used not only for the discretization of the interface (membrane) but also for the membrane mechanics solver (FEM -- the membrane force density) and the fluid dynamics solver (BEM -- the interfacial velocity and the disturbed wall traction).
In this way, the proposed numerical framework is constructed consistently with the IGA paradigm. Compared to the unbounded fluid-flow calculations, fluidic confinement raises the considerable difficulty in dealing with the hydrodynamic interaction between the wall boundary and the particle surface, as the bounded fluid-flow computations require highly refined grids to accurately capture wall-particle interactions.
Such a hydrodynamical computation frequently encounters numerical instabilities, particularly in the simulation of vesicle dynamics, which are associated with the membrane's bending rigidity and incompressibility, as reported in, e.g., Ref.~\cite{barakat_shaqfeh_2018b}. Thanks to the use of Loop elements in a consistent way, the present computational framework overcomes these challenges, enabling accurate and long-time numerically stable simulations. 

The rest of the paper is organized as follows. In Section~\ref{Sec:Form}, we describe the equations governing the flow motion and the membrane mechanics along with the interfacial conditions. In Section~\ref{Sec:Num}, we introduce Loop's subdivision scheme and discuss the numerical method implemented in the membrane and fluid solvers, whose validation is provided in Section~\ref{Sec:Val}. While the current numerical model can deal with channels of arbitrary cross-sections, we focus herein on a single deformable particle in a circular or rectangular channel, for demonstration purposes only. We present additional simulation results to demonstrate the accuracy and stability of the method in Section~\ref{Sec:Exp}, followed by conclusions and future directions in Section~\ref{Sec:Concl}.

\section{Problem statement and formulation}
\label{Sec:Form}
\begin{figure}[!htbp]
\centering\includegraphics[width=1.0\linewidth]{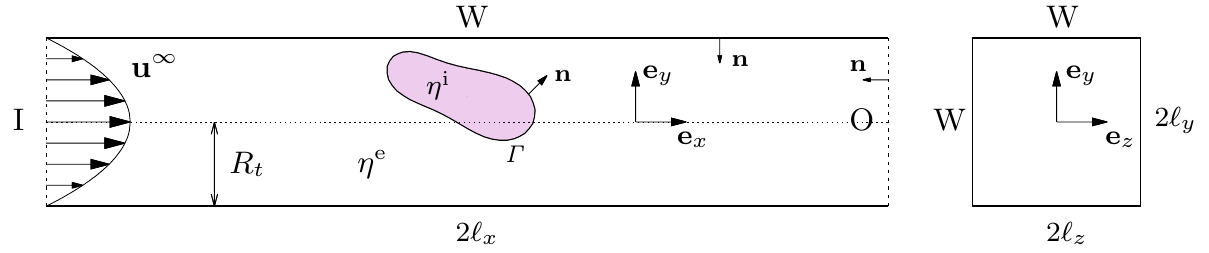}
\caption{Schematic representation of a freely suspended deformable particle (defined by the boundary $\Gamma$) transported in a pressure-driven flow through a straight microchannel (either circular of radius $R_t$ or rectangular of $2\ell_y$ by $2\ell_z$ in cross-section), with $\mathrm{I}$, $\mathrm{O}$ and $\mathrm{W}$ denoting the inlet, outlet and wall boundaries, respectively.  $\eta^\mathrm{i,e}$ are the viscosities of the fluids and $\bm{n}$ is the unit normal vector pointing into the suspending fluid.}
\label{fig:schematicFlow}
\end{figure}

As sketched in Fig.~\ref{fig:schematicFlow}, we consider a soft particle freely transported in a microchannel of constant cross-section. The particle deforms in response to the flow stresses of bulk flows, as well as to the wall boundary-induced viscous friction, resulting in a change in the membrane forces (e.g., the bending force), which in turn alter the bulk flows. The modeling of such an FSI involves formulating the following three parts: 

\begin{itemize}
 \item hydrodynamics;
 \item membrane mechanics;
 \item coupling conditions at the interface.
\end{itemize}

\subsection{Hydrodynamics}
The motion of the internal (with superscript i) and external (with superscript e) fluids is governed by the incompressible Newtonian Navier-Stokes equations.
In a wide variety of microfluidic flows, the Reynolds number is small (less than unity)~\cite{Stone_2004}. For example, in a channel of height \SI{100}{\um}, 
a flow with water (shear viscosity $\eta\approx 10^{-3}$~\si{\pascal\second}) at a typical speed \SI{1}{\mm\per\second} has the Reynolds number $Re \approx 0.1$.
The governing equations can therefore be reduced to the Stokes equations for creeping flow \par
\begin{equation}
-\bm{\nabla} p^{\mathrm{i,e}}+\eta^{\mathrm{i,e}}\nabla^2\bm{u}^{\mathrm{i,e}}=\bm{0},\quad \bm{\nabla} \cdot \bm{u}^{\mathrm{i,e}}=0,
\label{eq:StokesEquations}
\end{equation}
where $\bm{u}$ and $p$ denote the fluid velocity and pressure, respectively. The external velocity field $\bm{u}^{\mathrm{e}}$ satisfies the no-slip condition on the boundary walls of the channel  \par
\begin{equation}
\bm{u}^\mathrm{e}(\bm{x}) = \bm{0}, \qquad  \forall \bm{x} \in \mathrm{W}
\label{eq:solidWallBCs}
\end{equation}
and vanishing far-field flow perturbations, \par
\begin{equation} 
 \bm{u}^\mathrm{e}(\bm{x}) =   \bm{u}^{\infty}(\bm{x}), \qquad \forall \bm{x} \in \mathrm{I} \cup \mathrm{O} .
\label{eq:openSpaceBCs}
\end{equation}

\subsection{Membrane mechanics}
\label{sec:membraneMechanics}
A membrane is a closed and deformable interface separating the internal and external fluids. It is described by its position $\bm{x}(t)$ at time $t$. Under stresses, energy variation $\delta\mathit{E}$ is stored in the membrane through the elastic deformations (bending, shearing, and dilation/compression) or dissipated by viscous friction~\cite{Vlahovska_2013}.

The surface force density exerted by the membrane  $\bm{f}^m$ onto surrounding fluids is given by the first variation of its surface energy \par
\begin{equation}
 \bm{f}^m(\bm{x})=-\frac{1}{\sqrt{a}}\frac{\delta\mathit{E}}{\delta \bm{x}},\quad \mathit{E}=\int_{\Gamma}w_s\mathrm{d}S,
 \label{eq:forceDensity_energVariation}
\end{equation}
where $w_s$ is the surface energy per unit area which completely determines the mechanical properties of the membrane, and $a$ is the determinant of the local metric. Below we describe different formulations for $w_s$ (or $\bm{f}^m$), depending on the type of soft objects under study. 

For liquid drops, the surface energy per unit area is simply the interfacial tension $\gamma$ (which is a material property), i.e., $w_s=\gamma$, leading to  \par
\begin{equation}
 \bm{f}^m = \bm{\nabla}_s\gamma - 2\gamma \mathit{H}\bm{n},
 \label{eq:forceDensityForDrop}
\end{equation}
where $\bm{\nabla}_s=(\bm{I}-\bm{n}\bm{n}) \cdot \bm{\nabla}$ is the surface gradient operator with $\bm{I}$ the identity tensor, $\bm{n}$ is the outward pointing normal vector, and 
$\mathit{H}$ is the local mean curvature (with the convention that $\mathit{H}$ is positive for a sphere).

For vesicles, the lipid membrane is modeled as a two-dimensional incompressible fluid with bending stiffness.
The local surface-area incompressibility of the membrane    \par
\begin{equation}
\bm{\nabla}_s \cdot \bm{u}=0
\label{eq:incomp}
\end{equation}
 is enforced via the Lagrange field $\gamma$ (equivalent to the membrane tension, which is not a material property but is  flow dependent). This  membrane tension is added to the bending energy density $w^{\mathit{H}}$~\cite{Helfrich_1973}, giving the elastic energy density of a vesicle\par
\begin{equation}
 w_s=w^{\mathit{H}}_s+\gamma,\quad w^{\mathit{H}}_s=\frac{\kappa}{2}\left(2\mathit{H}\right)^2,
 \label{eq:energyDensity_vesicle}
\end{equation}
where $\kappa$ is the bending modulus of the lipid bilayer. In principle, a spontaneous (or reference) curvature and Gaussian curvature ($\mathit{K}$) appear in the bending energy. For simplicity, we take the minimum energy reference state as a flat sheet. The term with Gaussian curvature does not contribute to
variation of the bending energy if the topology remains unchanged, which is the case of our study. Using Eq.~\eqref{eq:forceDensity_energVariation}, one obtains a formal expression of the surface force density~\cite{Zhong_can_1989} \par
\begin{equation}
\bm{f}^m=\kappa\left[2\Delta_s\mathit{H}+4\mathit{H}(\mathit{H}^2-\mathit{K})\right]\bm{n}+\bm{\nabla}_s\gamma -2\gamma \mathit{H}\bm{n},
\label{eq:forceDensityForVesicle}
\end{equation}
where $\Delta_s=\bm{\nabla}_s \cdot \bm{\nabla}_s$ is the Laplace-Beltrami operator, which contains the fourth derivative of the surface position, posing numerical challenges to compute the bending forces~\cite{Guckenberger_CPC_2016}.

For capsules with a vey thin hyperelastic membrane, two popular membrane constitutive laws are used herein:  the Neo-Hookean (NH) law and
the Skalak (Sk) law~\cite{Barthes_2016}.
For these laws, the surface density of membrane energy is defined upon a reference configuration $S^0$ as, \par
\begin{equation}
 \left\{ 
 \begin{aligned}
  w^{NH}_{s^0} &= \frac{\mu_s}{2}\left[I_1-1+\frac{1}{I_2+1}\right] \\
  w^{Sk}_{s^0} &= \frac{\mu_s}{4}\left[I^2_1+2I_1-2I_2+CI^2_2\right]
 \end{aligned}
 \right. ,
 \label{eq:energyDensityForCapsule}
\end{equation}
where $\mu_s$ is the surface shear modulus, $C$ represents the relative importance of the resistance to surface dilation, and $I_1$ and $I_2$ are the two strain invariants.  \par

Finally, for red blood cells having a composite membrane, we use our recent RBC membrane model~\cite{Lyu_2018} to compute the surface force density. Briefly, the RBC membrane is modeled as a composite network, which consists of a dynamically triangulated surface as in a fluid vesicle model. The membrane is then coupled to an additional network of springs with fixed connectivity, representing the cytoskeleton. We explicitly compute the mechanical interaction between the bilayer and the cytoskeleton by considering normal elastic spring and tangential friction force.
Specifically, the FENE-POW spring model is used to describe the elastic cytoskeleton, which yields a spring force at node $n$ by an edge $np$ \par
\begin{equation}
 \mathbf{f}^{e}=\mathbf{f}_{np} = \frac{4\mu_\mathrm{s}}{\sqrt{3} \left ( \frac{2x_0^2}{1-x_0^2} + \alpha +1 \right )}
\left ( \frac{1-x_0^2}{1-x_{np}^2} - \frac{x_0^{\alpha+1}}{x_{np}^{\alpha+1}} \right) (\bm{x}_p - \bm{x}_n),
\label{eq:edgeSpringForce}
\end{equation}
where $\bm{x}_p$ ($\bm{x}_n$) is the position of vertex $p$ ($n$). The normalized spring length $x_{np}=l_{np}/l^{\max}_{np}\in \left(0, 1\right]$ (the ratio of the spring length and maximum spring length),
$x_0=x^0_{np}$ denotes the  normalized spring length of edge $np$ in the reference shape, and $\alpha$ is a constant repulsive parameter.
This force, embodied on each spring edge, is transmitted to the lipid bilayer in the normal direction directly and in the tangential plane indirectly via drag forces. In this way, interfacial viscosity is added (see Ref.~\cite{Lyu_2018} for details).

\subsection{Coupling conditions}
\label{sec:couplingConditions}
The interface conditions need to be imposed to complete the problem formulation. First, the fluid motion is coupled with the interface motion via the kinematic boundary condition, i.e., continuity of the velocities at the interface \par
\begin{equation}
 \bm{u}^\mathrm{e}(\bm{x})=\bm{u}^\mathrm{i}(\bm{x})=\bm{u}_\Gamma,\quad \forall\bm{x}\in\Gamma ,
 \label{eq:velocityContinuityAtTheInterface}
\end{equation}
where $\bm{u}_\Gamma$ is the velocity of fluids at the interface. 

Second, assuming an impermeable membrane, at least on typical experimental time scales, the membrane is advected by the interface flow \par
\begin{equation}
\frac{\mathrm{d}\bm{x}}{\mathrm{d} t}=\bm{u}_\Gamma,\quad \forall\bm{x}\in\Gamma ,
 \label{eq:velocityContinuityOfFluidAndMembrane}
\end{equation}
where $\bm{x}$ is the membrane position. 

Finally, the dynamic boundary condition at the interface establishes a nonlinear interaction between bulk flows and 
membrane mechanics, \par
\begin{equation}
\Delta \bm{f} +\bm{f}^m=\bm{0},
 \label{eq:forceBalanceAtTheMembrane}
\end{equation}
wherein we assume the membrane is in quasi-static mechanical equilibrium; the membrane force density $\bm{f}^m$ balances the traction jump $\Delta \bm{f}$ ($\equiv (\bm{\sigma}^\mathrm{e}-\bm{\sigma}^\mathrm{i}) \cdot  \bm{n}$)  exerted on the membrane by bulk fluids,
with the stress tensor $\bm{\sigma} \equiv -p\bm{I} + \eta\left[\bm{\nabla u}+(\bm{\nabla u})^T\right]$.

We also compute several qualities of interest to show the simulation results, such as the particle's shape and mobility. The translational velocity of the particle's center of mass in the streamwise direction is given by \par
\begin{equation} 
U_x =  \frac{1}{V}  \int_V (\bm{u}^\mathrm{i} \cdot \bm{e}_x)\,\mathrm{d}^3\bm{x} = \frac{1}{V} \int_\Gamma x (\bm{u}_\Gamma \cdot \bm{n})\,\mathrm{d}S(\bm{x}) ,
 \label{eq:Vel}
\end{equation}
where $V$ is the enclosed volume of the particle, which is calculated from \par
\begin{equation} 
V =   \int_V \mathrm{d}^3\bm{x} = \frac{1}{3} \int_\Gamma (\bm{x} \cdot \bm{n})\mathrm{d}S(\bm{x}).
 \label{eq:Vol}
 \end{equation}
 Its derivation from the initial given volume during simulations provides an indication of the accuracy of the computations. For vesicles, the relative surface area variation is also an indicator of the accuracy. The coordinates of the particle's center of mass are given by \par
 \begin{equation}
\label{eq:mass }
\bm{X}_g \cdot \bm{e}_i = \frac{1}{V}\int_V \bm{x} \,\mathrm{d}^3\bm{x} =  \frac{1}{2V}\int_\Gamma (\bm{x} \cdot \bm{e}_i)^2 (\bm{n} \cdot \bm{e}_i)  \,\mathrm{d}S(\bm{x}).
\end{equation}

\section{Numerical method} \label{Sec:Num}

We use Loop's subdivision elements~\cite{Loop_1987} to represent every quantity/field of interest: meshes (of the surface of objects and the wall surface of channels) and the unknown density fields -- the interfacial velocity, the membrane force, and the disturbed wall traction. We begin with Loop's subdivision scheme, followed by a description of how the membrane forces are calculated with a unified formalism. The thus-obtained membrane forces are then used to compute the interfacial velocity using Green's function. Finally, we describe the interface advection schemes. 

\subsection{Isogeometric analysis}
 
\subsubsection{Subdivision surfaces}
\begin{figure}[!htbp]
 \centering\includegraphics[width=0.9\linewidth]{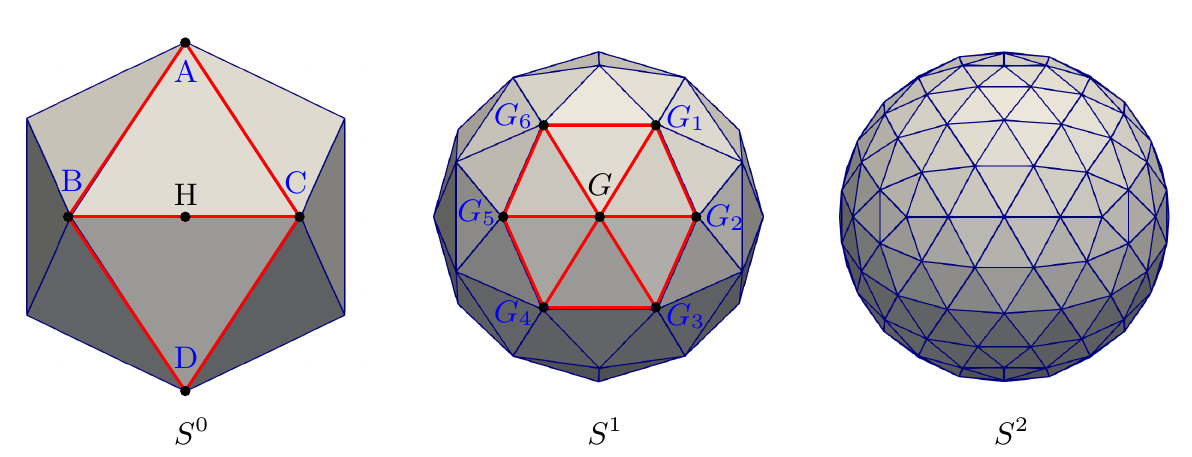}
 \caption{An illustration of Loop's subdivision rule from the initial control mesh $S^0$ (an icosahedron) to $S^1$ mesh (after one refinement) and $S^2$ mesh (after two refinements).}
 \label{fig:LoopSubdivisionRule}
\end{figure}

Loop's subdivision surface is an assembly of linear triangle elements generated through a limiting procedure of repeated refinement starting from an initial coarse mesh, called the control mesh of the surface. For continuously deformed particles, 
an icosahedron containing 20 equilateral triangle faces with five meeting at each of its 12 vertices can be used as the initial control mesh ($S^0$ in Fig.~\ref{fig:LoopSubdivisionRule}). The control mesh and all refined meshes (by quadrisection) consist of triangles only. 

In Loop's subdivision scheme, each triangle of the coarse mesh is quadrisected by introducing a new vertex at each edge midpoint, as illustrated in Fig.~\ref{fig:LoopSubdivisionRule}.
The coordinates of the newly generated vertices (level $k+1$) on the edge of the previous mesh (level $k$) are computed as \par
\begin{equation}
 p^{k+1}_H=\frac{p^k_A+3p^k_B+3p^k_C+p^k_D}{8},
 \label{eq:insertNewVertex}
\end{equation}
and the old vertices are updated to get new nodal positions at the mesh $k+1$ \par
\begin{equation}
 p^{k+1}_G=(1-q\varpi)p^k_G+\varpi\sum^n_{i=1}p^k_{G_i}.
 \label{eq:updateExistingVertex}
\end{equation}
Here $G_i$ ($i\in[1,q]$) are the one-ring neighbours (at level $k$) of the vertex $G$, i.e., those vertices which share an edge with it, and $q$ denotes the valence of a vertex, the number of element edges attached to a vertex~\cite{Cirak_2000}. \par
The value of $\varpi$, proposed by Loop~\cite{Loop_1987}, is given by \par
\begin{equation}
 \varpi=\frac{1}{q}\left[\frac{5}{8}-\left(\frac{3}{8}+\frac{1}{4}\cos\frac{2\pi}{q}\right)^2\right].
 \label{eq:parameterForLoopRefinement}
\end{equation}

Note that almost all newly generated vertices are regular (with valence $q=6$), except for
the twelve vertices (with valence $q=5$) updated from the initial icosahedron mesh, which remain irregular. 
Loop's subdivision scheme produces limit surfaces which are globally $C^2$ except at those irregular points where they are only $C^1$. However, the surfaces obtained by this scheme are $H^2$, i.e., have finite bending energy.
\begin{figure}[!htbp]
 \centering\includegraphics[width=0.9\linewidth]{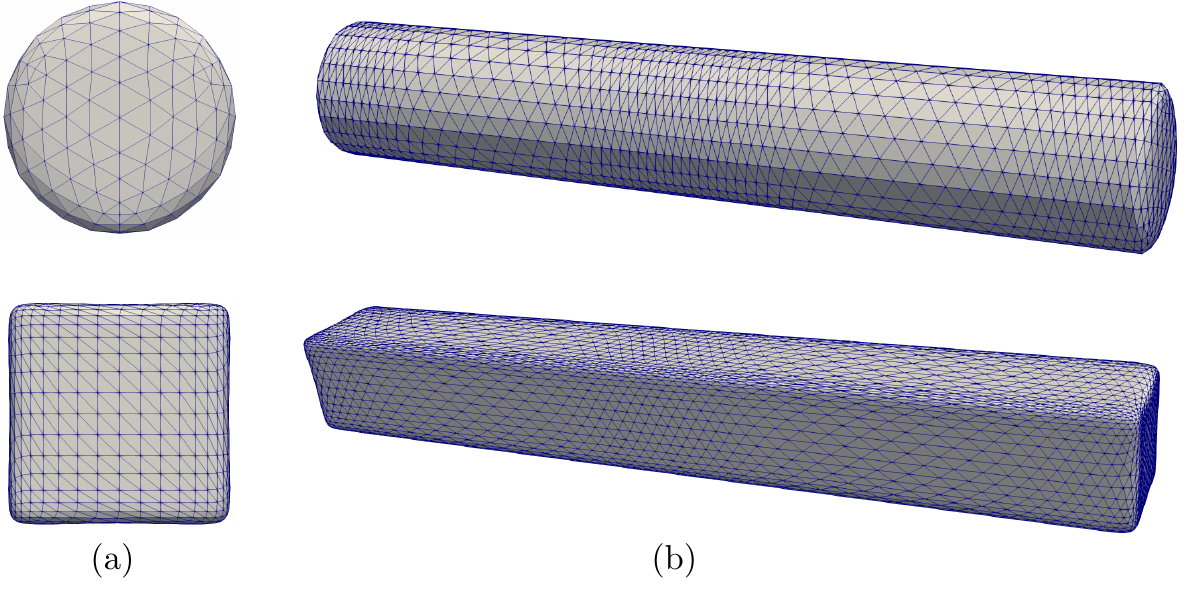}
  \caption{An example of closed wall meshes generated by Loop's subdivision scheme. (a) on the inlet and outlet (circular and square) sections and (b) on the wall surface of (circular and square) microchannels. The mesh comprises $N=3360$ elements and  1682 nodes for the circular tube and $N^w=7427$ elements and 3714 nodes for the square channel. }
 \label{fig:tubeMeshExamples}
\end{figure}

Two typical microchannel meshes generated by Loop subdivision are shown in Fig.~\ref{fig:tubeMeshExamples}, one for cylindrical channel and the other for a square channel.
Since the soft particle is kept at the center of the channel, this region has a more refined mesh on the wall surface.
The intersections (of the wall and the inlet/outlet surfaces) and the corners of the square channel are rounded with an arc-circle (radius $\in \left[O (R_t/6), O (R_t/4)\right]$ 
to avoid corner effects when solving the flow with the boundary element method~\cite{Pozrikidis_1992,Hu_JFM_2012}.
\subsubsection{Isogeometric representation}

 \begin{figure}[!htbp]
 \centering
    \includegraphics[width=0.4\textwidth]{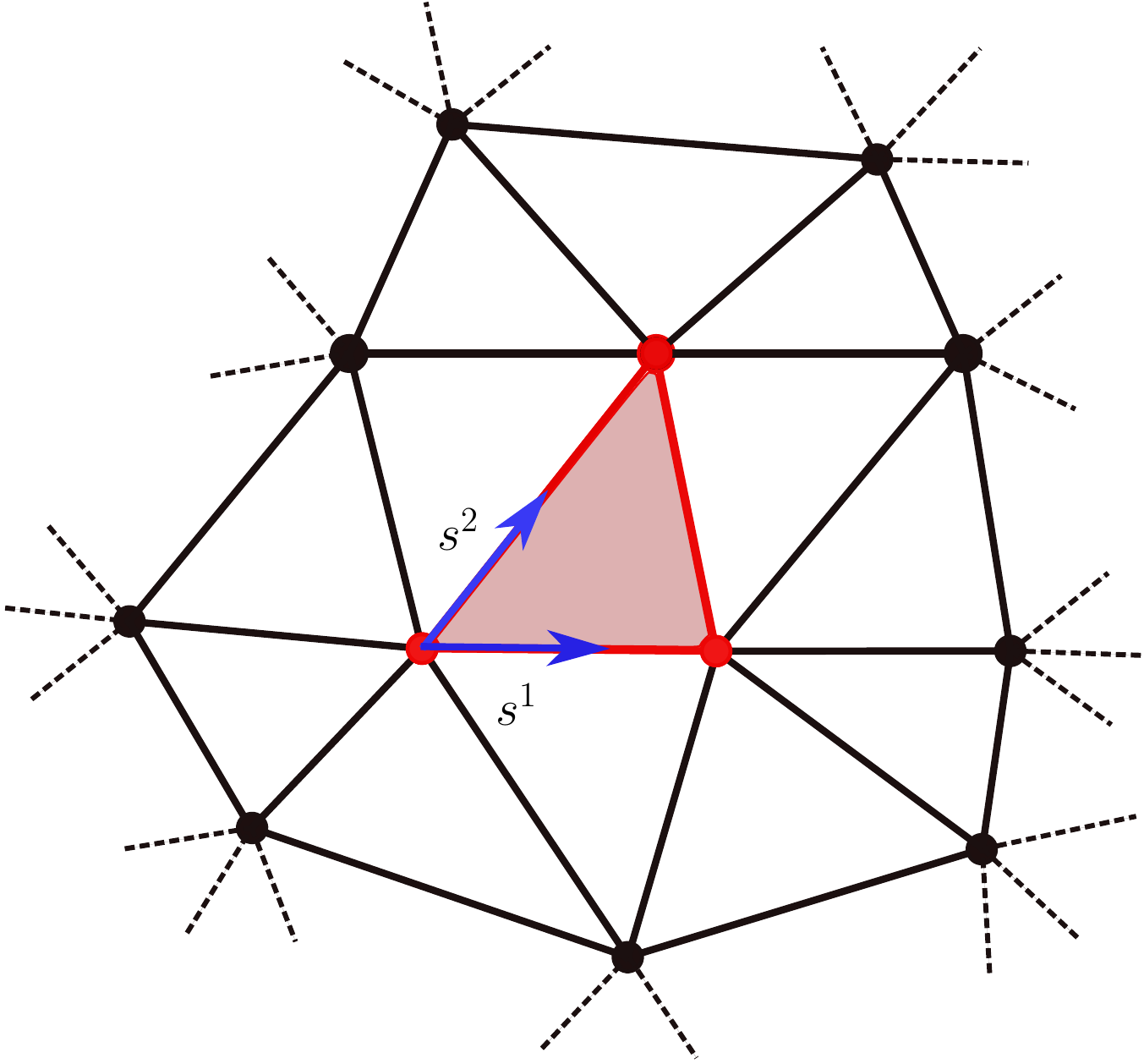}
 \caption{A Loop element (shaded triangle) with its local parametrization $\bm{s}=(s^1,s^2)$ and its one-ring elements (bounded by solid lines), forming a regular Loop's patch containing 12 control vertices.}
 \label{fig:subdivisionOneRingShapeFunc}
 \end{figure}

Stam~\cite{Stam_1998} shows that the limit position of any point inside a triangle element $e$ may be expressed in terms of box-spline shape functions, \par
\begin{equation}
 \bm{x}^e=\sum_{p\in\text{one-ring}}\bm{X}^p N_p (s^1,s^2),
 \label{eq:limitPositionLoop}
\end{equation}
where the sum is taken over one-ring vertices, as shown in Fig.~\ref{fig:subdivisionOneRingShapeFunc}, $(s^1,s^2)$ is a local parametrization of this point on the element, 
and $N_p$ are the shape functions spanning over all one-ring elements~\cite{Cirak_2000,Stam_1998}.
The nodal values $\bm{X}^p$ are the approximation parameters of the limit position $\bm{x}^e$ in the space expanded by the shape functions $N_p$. The parameterization in Eq.~(\ref{eq:limitPositionLoop}) may also be used for any scalar function $f$ defined on the membrane or at the channel wall, e.g., a Cartesian component of the membrane force $\bm{f}^m$, the interfacial velocity $\bm{u}$, and  the disturbed wall traction $\bm{f}^w$,  \par
\begin{equation}
 f^e(\bm{x})=f^e(s^1,s^2)=\sum_{p\in\text{one-ring}}F^p N_p(s^1,s^2),
 \label{eq:fieldExpansionLoop}
\end{equation}
where $F^p$ is the $p$-th nodal value. Equation~\eqref{eq:fieldExpansionLoop} is used to evaluate $f$ (i.e., the limit value) at any position $\bm{x}$ on element $e$ if the nodal values $F^p$ are known. 
Inversely, we also need to convert the limit value of a field $f$ into its nodal values $F^p$, that is, given the approximation of $f$ under the form~\eqref{eq:fieldExpansionLoop} such that the approximation error is minimized. Using the collocation formulation, in which the known field $f$ is collocated at vertices, i.e., $f^n=f(\bm{x}=\bm{x}^n)$ is known at vertex $\bm{x}^n$ \par

\begin{equation}
f^n=\sum_{p\in\text{one-ring}}F^pN_p(s^1(\bm{x}^n),s^2(\bm{x}^n))\quad \forall n\in \{1,\cdots,N_\mathrm{v}\},
% f_p=\sum_{q\in\text{one-ring}}F_qN^q(s^1(\bm{x}_q),s^2(\bm{x}_q)),
\label{eq:collocationSubdivision}
\end{equation}
where $N_p(s^1(\bm{x}^n),s^2(\bm{x}^n))$ are the shape functions evaluated in the local parameter space $(s^1,s^2)$ corresponding to the vertex $\bm{x}^n$, and $N_\mathrm{v}$ is the total number of vertices. 
Assembling the linear system~\eqref{eq:collocationSubdivision} in matrix form according to the index of vertices, we then have \par
\begin{equation}
 \{f^n\}=\mathbf{C}\{F^n\},
 \label{eq:collocationSubdivision2}
\end{equation}
where $\{f^n\}=\{f^1,~f^2,~\cdots ~,f^{N_\mathrm{v}}\}^T$, $\{F^n\}=\{F^1,~F^2,~\cdots ~,F^{N_\mathrm{v}}\}^T$, and $\mathbf{C}$ is the collocation matrix that transforms between
the limit values $f^n$ and the nodal values $F^n$.

The regular surface patch is a quadratic spline~\cite{Stam_1998}, the derivatives of first- and second-order can thus be realized by directly deriving on the shape functions, such as \par
\begin{equation}
 f^e_{,\alpha}(s^1,s^2)=\sum_{p\in\text{one-ring}}F^pN_{p,\alpha}(s^1,s^2), 
 \label{eq:derivationOnRegularElements}
\end{equation}
where here and henceforth Greek indices takes the values 1 and 2, and a comma is used to denote partial differentiation.

For irregular Loop elements, the irregular patch must be subdivided until the parameter value $(s^1, s^2)$ of interest is within a regular patch, 
and then the canonical regular-patch evaluation routine works again~\cite{Cirak_2000,Cirak_2011}.

\subsection{Membrane solver -- FEM}
\label{sec:numMembraneSolver}
\subsubsection{Weak formulation of FSI}
The membrane solver is designed to calculate the membrane force density $\bm{f}^m$ using the finite element method. The solver is based on the principle of virtual work for a deformable body. A detailed description is provided in~\cite{Boedec_JCP_2017}. For the sake of completeness, we recall briefly some basic concepts below.  

 To describe the deformation of a surface from a reference configuration $\bm{x}^0(\bm{s})$ to its current configuration $\bm{x}(\bm{s})$, we introduce the tangential vectors at a point on the surface, which are given by the covariant base vectors $\bm{a}_\alpha$  \par
\begin{equation}
 \bm{a}_\alpha (\bm{s}) =\frac{\partial\bm{x}(\bm{s})}{\partial s^\alpha}=\bm{x}_{,\alpha}(\bm{s}).
 \label{eq:localSurfaceBasis}
\end{equation}
The unit normal vector $\bm{n}$ can be written as \par
\begin{equation}
 \bm{n}(\bm{s})=\frac{\bm{a}_1(\bm{s}) \times\bm{a}_2(\bm{s})}{|\bm{a}_1(\bm{s})\times\bm{a}_2(\bm{s})|}.
 \label{eq:localSurfaceNorm}
\end{equation}
Contravariant base vectors $\bm{a}^\alpha$ are obtained through the relation $\bm{a}^\alpha \cdotp \bm{a}_\beta = \delta^\alpha_\beta$, where $\delta^\alpha_\beta$ is the Kronecker delta. 

With the tangential and normal vectors, we can write the first and second fundamental forms of the surface \par
\begin{equation}
 a_{\alpha\beta}(\bm{s})=\bm{a}_\alpha(\bm{s})\cdotp\bm{a}_\beta(\bm{s}),\quad b_{\alpha\beta}(\bm{s})=\bm{a}_{\alpha,\beta}(\bm{s})\cdotp\bm{n}(\bm{s}),
 \label{eq:metricCurvatureTensors}
\end{equation}
where $a_{\alpha\beta}$ and $b_{\alpha\beta}$ denote the metric and curvature tensors of the surface, respectively. The differential area element of the surface $\mathrm{d}S=\sqrt{a}\mathrm{d}s^1\mathrm{d}s^2$, with $a=\det (a_{\alpha\beta})$ the determinant of the metric tensor. The above definitions and relations hold for the reference configuration as well, with $\bm{x}^0(\bm{s})$ replacing $\bm{x}(\bm{s})$.

By virtue of the principle of virtual work, a membrane is in equilibrium if the sum of internal and external virtual work vanishes \par
\begin{equation}
\delta W_\mathrm{int} + \delta W_\mathrm{ext} =0.
\label{eq:virtualWorkForShell}
\end{equation}
The external virtual work is given by \par
\begin{equation}
\delta W_\mathrm{ext} =\int_\Gamma  \left (\Delta \bm{f} + \bm{g} \right ) \cdotp\delta\bm{x}\mathrm{d}S,
\end{equation}
where (and in what follows) $\delta$ means that a variable derives from a virtual displacement $\delta \bm{x}$, and $\bm{g}$ is some additional body forces (e.g., buoyancy) acting on the membrane in addition to the term $\Delta \bm{f}$ representing the traction jump across the membrane.

According to Ref.~\cite{Cirak_2000}, the internal virtual work of the membrane can be written as \par
\begin{equation}
 \delta W_\mathrm{int} = -\int_\Gamma \left[\sigma^{\alpha\beta}\delta(E_{\alpha\beta})+\mu^{\alpha\beta}\delta(B_{\alpha\beta})\right]\mathrm{d}S,
\end{equation}
where $\sigma^{\alpha\beta}$ and $\mu^{\alpha\beta}$ are the effective membrane and bending stress tensors, respectively. They are membrane dependent -- namely its position and mechanical properties.  The Green-Lagrange strain tensor \par
\begin{equation}
 E_{\alpha\beta}=\frac{1}{2}\left(a_{\alpha\beta}-a^0_{\alpha\beta}\right)
 \label{eq:stretchingStrainTensor}
\end{equation}
represents in-plane deformation, i.e., stretching, while the bending strain tensor \par
\begin{equation}
 B_{\alpha\beta}=b_{\alpha\beta}-b^0_{\alpha\beta}
 \label{eq:bendingStrainTensor}
\end{equation}
describes out-of-plane deformation, i.e., the change in curvature or bending strains.

Note that $\Delta \bm{f} =-\bm{f}^m$ (Eq.~(\ref{eq:forceBalanceAtTheMembrane})), so for neutrally buoyant particles, Eq.~\eqref{eq:virtualWorkForShell} reads \par
\begin{equation}
 \int_\Gamma \left[\frac{1}{2}\sigma^{\alpha\beta}\delta(a_{\alpha\beta})+\mu^{\alpha\beta}\delta(b_{\alpha\beta}) + \bm{f}^m\cdotp\delta\bm{x}\right]\mathrm{d}S=0, \quad \forall \delta \bm{x} \in H^2(\Gamma). 
 \label{eq:virtualWorkForShell2}
\end{equation}
This weak formulation of a fluid-membrane interaction problem gives a general relationship between membrane force density $\bm{f}^m$ and its position $\bm{x}$. Using isogeometric finite element which ensures $H^2$, this unified formalism makes it possible to study deformable objects spanning from a simple liquid drop to elastic capsules and vesicles with bending stiffness. The membrane forces are obtained once the mechanical properties of the membrane are specified via the membrane ($\sigma^{\alpha\beta}$) and bending ($\mu^{\alpha\beta}$) stress tensors. 

\subsubsection{Membrane constitutive laws}
For liquid drops, the surface energy per unit area is the interfacial tension, $w_s=\gamma$, independent of curvature, thus the bending stress tensor $\mu^{\alpha\beta}=0$. Hence, the membrane stress tensor is given by \par
\begin{equation}
 \sigma^{\alpha\beta}=\gamma a^{\alpha\beta}.
 \label{eq:membraneStressesForDrop}
\end{equation}

For a capsule with a very thin elastic membrane, the surface energy density $w_s$ is given by Eq.~\eqref{eq:energyDensityForCapsule} for the NH and Sk laws, we also have  $\mu^{\alpha\beta}=0$. Following~\cite{Walter_2010,Maestre_CMAME_2017}, the membrane stress tensor can be written as \par
\begin{equation}
 \sigma^{\alpha\beta}=\frac{2}{J_s}\frac{\partial w_s}{\partial I_1}a^{0,\alpha\beta}+2J_s\frac{\partial w_s}{\partial I_2}a^{\alpha\beta},
 \label{eq:membraneStressesForCapsule}
\end{equation}
where $J_s=\sqrt{a}/\sqrt{a^0}$ is the Jacobian of the transformation from the reference to the deformed configuration.

For a lipid membrane satisfying the Helfrich bending energy subjected to the surface incompressibility constraint  (Eq.~\eqref{eq:forceDensityForVesicle}), the membrane and 
bending stress tensors are given by~\cite{Boedec_JCP_2017} \par
\begin{equation}
 \left\{ 
 \begin{aligned}
  \sigma^{\alpha\beta} &= \frac{2}{\sqrt{a}}\frac{\partial\left(\sqrt{a}w^H_s\right)}{\partial a_{\alpha\beta}}=\frac{\kappa}{2}\left(4H^2 a^{\alpha\beta}-8Hb^{\alpha\beta}\right)+\gamma a^{\alpha\beta} \\
  \mu^{\alpha\beta} &= \frac{\partial w^H_s}{\partial b_{\alpha\beta}}=\frac{\kappa}{2}\left(4Ha^{\alpha\beta}\right)
 \end{aligned}
 \right. .
 \label{eq:membraneStressesForVesicle}
\end{equation}

Finally, for an RBC having a composite membrane, the cytoskeletal elastic forces $\bm{f}^{e}$ are computed directly based on a spring network (Eq.~(\ref{eq:edgeSpringForce})). These forces are added to the lipid bilayer (Eqs.~(\ref{eq:membraneStressesForVesicle}) and (\ref{eq:virtualWorkForShell2}) in the normal direction directly  and in the tangential plane indirectly via drag forces~\cite{Lyu_2018}.

\subsubsection{Calculation of membrane force}

Since surfaces obtained by Loop's subdivision scheme are globally $C^2$ except at some fixed irregular points, the curvature tensor $b_{\alpha\beta}$ at any quadrature points (12 Gauss quadrature points are used) can be computed by direct differentiation of Loop's shape functions. The metric tensor $a_{\alpha\beta}$ and the unit normal vector $\bm{n}$ are readily obtained from the interpolation of the position. As such, the finite-element discretization of Eq.~(\ref{eq:virtualWorkForShell2}) using the Loop shape functions for the Cartesian components of membrane force and position leads to a matrix-vector form for the unknown nodal values of the membrane force $\mathbf{f}^m$  \par
\begin{equation}
\mathbf{M}\{\mathbf{f}^m\}=\{\mathbf{rhs}\}.
\label{eq:virtualWorkForShell2_matrixForm}
\end{equation}
The mass matrix $\mathbf{M}$ and the right hand side vector $\{\mathbf{rhs}\}$ are formed by numerical integration of Eq.~(\ref{eq:virtualWorkForShell2}) using 12 Gauss quadrature points, see Ref.~~\cite{Boedec_JCP_2017} for details.

\subsection{Fluid solver -- BEM}
Under Stokes flow conditions, boundary integral equations for the interfacial velocity and the disturbed wall traction can be expressed as~\cite{Pozrikidis_1992,Pozrikidis_2005}, on the membrane surface 
\begin{eqnarray}
\frac{1+\lambda}{2}\bm{u}(\bm{x}_0) & = & \bm{u}^\infty(\bm{x}_0) + \mathcal{S}_{\Gamma} \bm{f}^m(\bm{x}_0) - \mathcal{S}_\mathrm{W}\bm{f}^w(\bm{x}_0) 
-\Delta p^a\mathcal{S}_\mathrm{O}\bm{n}(\bm{x}_0) \nonumber\\
& & + (1-\lambda)\mathcal{D}^{PV}_{\Gamma}\bm{u}(\bm{x}_0), \quad \bm{x}_0 \in\Gamma 
\label{eq:u_interface}
\end{eqnarray}
and at the channel wall (the no-slip condition) \par
\begin{equation}
\mathcal{S}_\mathrm{W}\bm{f}^w(\bm{x}_0) = \mathcal{S}_{\Gamma} \bm{f}^m(\bm{x}_0)  -\Delta p^a\mathcal{S}_\mathrm{O} \bm{n}(\bm{x}_0)
 + (1-\lambda)\mathcal{D}_{\Gamma}\bm{u}(\bm{x}_0), \quad \bm{x}_0 \in \mathrm{W} 
\label{eq:wall_stress}
\end{equation} 
where $\bm{u}$ is the interfacial velocity, $\bm{u}^\infty$ is the velocity of the ambient flow (i.e., the flow without the deformable object), $\bm{f}^w$ is the disturbed wall traction, and $\lambda$ ($\equiv \eta^{\mathrm{i}}/\eta^{\mathrm{e}}$) is the viscosity ratio between the internal and external fluids. The single-layer operator $\mathcal{S}$ and double-layer operator $\mathcal{D}$ are defined as \par
\begin{subequations} \label{E:SD}
  \begin{gather}
 \left( \mathcal{S}_{\Omega}\bm{\psi}\right)_j(\bm{x}_0)=\frac{1}{8\pi\eta^\mathrm{e}}\int_\Omega \psi_i(\bm{x})G_{ij}(\bm{x},\bm{x}_0)\mathrm{d}S(\bm{x}),  \label{E:SD1} \\
  \left( \mathcal{D}_{\Omega}\bm{\psi}\right)_j(\bm{x}_0)= \frac{1}{8\pi}\int_\Omega  \psi_i(\bm{x})T_{ijk}(\bm{x},\bm{x}_0)n_k(\bm{x}) \mathrm{d}S(\bm{x}),  \label{E:SD2}
      \end{gather}
\end{subequations}
where $\bm{G}$ (Stokeslet) and $\bm{T}$ (stresslet)  are the Green's functions in the three-dimensional free space~\cite{Pozrikidis_1992}. $\mathcal{D}^{PV}$ indicates  that the double-layer integral is evaluated in the principal-value sense when the point $\bm{x}_0$ lies on the integration domain $\Gamma$. The term $\Delta p^{a}$, called additional pressure drop, is due to the presence of a particle in the channel flow which causes an increase in the pressure drop across the channel. It can be calculated by the reciprocal theorem of Stokes flow~\cite{Pozrikidis_2005} \par
\begin{equation}
\Delta p^a=-\frac{1}{Q} \int_\Gamma \left [ \bm{f^m} \cdot \bm{u}^\infty+ (1 - \lambda) \bm{f}^\infty \cdot \bm{u} \right ]\mathrm{d}S(\bm{x}), \quad \bm{x} \in\Gamma
\label{eq:disturbancePressureDrop_equalVisco}
\end{equation}
where $Q$ is the total flow rate, which is assumed not disturbed by the presence of the deformable object, i.e., $Q=Q^\infty$.

The equations (\ref{eq:u_interface}) and (\ref{eq:wall_stress}), together with (\ref{eq:disturbancePressureDrop_equalVisco}) allow to determine the interfacial velocity $\bm{u}$ and the disturbed wall traction $\bm{f}^w$, as well as the additional pressure drop $\Delta p^{a}$. The last quantity has a direct implication in the rheological properties of a dilute suspension~\cite{Pozrikidis_2005}. For the sake of simplicity, we performed computations in this paper only with unity viscosity ratio (i.e., $\lambda=1$), unless specified otherwise.  The singularity in the single-layer integrals is treated in two ways, depending on the integration domain; one consists in a singularity subtraction technique proposed in \cite{Farutin_JCP_2014} when the integration domain lies on the membrane surface, and the other involves transforming the parametric triangular to polar coordinates as introduced in \cite{Pozrikidis_1992} when it lies at the channel wall. 
%Non-singular integrals are calculated using a 12 point Gaussian quadrature formulation for triangular elements.

For a vesicle, an additional field, namely the membrane tension $\gamma$, remains to be determined. It is the solution of the surface velocity incompressibility constraint (\ref{eq:incomp}), which is solved by an iterative method~\cite{Boedec_JCP_2017}.

\subsection{Time-stepping schemes}
\label{sec:timeStepping}
As have been implemented in~\cite{Boedec_JCP_2017}, two time-stepping schemes, i.e., a high-order explicit scheme and a second-order implicit scheme, are used to update the new membrane position $\bm{x}_{n+1}$ at time $t_{n+1}=t_n+\Delta t$.

\subsubsection{Runge-Kutta-Fehlberg scheme}
The explicit time-stepping scheme consists of a Runge-Kutta Fehlberg fourth-fifth (RKF45) stage scheme~\cite{Fehlberg_1969}. This high-order scheme allows dynamically adapting 
the time step $\Delta t=h_n$ and provides very good conservation of invariants such as the enclosed fluid volume.
The fourth and fifth stage formulations are given by \par
\begin{equation}
\bm{x}^{(4)}_{n+1}=\bm{x}_n+\frac{25}{216}\bm{k}_1 + \frac{1408}{2565}\bm{k}_3 + \frac{2197}{4101}\bm{k}_4 - \frac{1}{5}\bm{k}_5
\label{eq:RKF_fourthStage}
\end{equation} 
\begin{equation}
\bm{x}^{(5)}_{n+1}=\bm{x}_n+\frac{16}{135}\bm{k}_1 + \frac{6656}{12858}\bm{k}_3 + \frac{28651}{56430}\bm{k}_4 - \frac{9}{50}\bm{k}_5+\frac{2}{55}\bm{k}_6,
\label{eq:RKF_fifthStage}
\end{equation} 
where $\bm{k}_i$ correspond to the intermediate values~\cite{Fehlberg_1969}. The dynamic time step is realized by comparing the difference between the fourth and fifth stage 
results $\epsilon_{n}=\max |\bm{x}^{(4)}_{n+1}-\bm{x}^{(5)}_{n+1}|$ with the two pre-setting tolerances $\varepsilon_{\max}$ ($\leq 10^{-7}$) and $\varepsilon_{\min}$ ($\leq 10^{-8}$) \par
\begin{equation}
\left\{
\begin{aligned}
h_n &= \left(\frac{\varepsilon_{\max}}{2\epsilon_n}\right)^{1/4} h_n\quad\quad \mathrm{if}~\epsilon_{n}>\varepsilon_{\max} \\
h_{n+1} &=  \left(\frac{\varepsilon_{\min}}{2\epsilon_n}\right)^{1/4} h_n\quad\quad \mathrm{if}~\epsilon_{n}<\varepsilon_{\min} \\
h_{n+1} &= h_n \hspace{3cm} \mathrm{else}
\end{aligned}
\right. .
\end{equation}

\subsubsection{Trapezoidal scheme}
While the RKF45 time scheme is usually used in the simulation of drops and capsules, the bending stiffness of a vesicle precludes its use in the simulation of vesicle dynamics since the stability condition imposes very small time-steps, namely $\Delta t\leq O(\eta^\mathrm{e}\Delta x^3/\kappa)$, for an explicit time-stepping scheme to be numerically stable~\cite{Zhao_JCP2010}. Hence, an implicit scheme is needed. 

The implicit time scheme is the trapezoidal rule -- an implicit second-order Crank-Nicolson time integration. For a given position and tension $(\bm{x}_n,\gamma_n)$ at time $t_n$, 
the position and tension  $(\bm{x}_{n+1},\gamma_{n+1})$ at $t_{n+1}$ are nonlinearly coupled such as \par
\begin{equation}
 \left\{
 \begin{aligned}
  &\bm{x}_{n+1}=\bm{x}_n+\frac{\Delta t}{2}\left[\bm{u}(\bm{x}_n,\gamma_n)+\bm{u}(\bm{x}_{n+1},\gamma_{n+1})\right] \\
  &\bm{\nabla}_s\cdotp\bm{u}(\bm{x}_{n+1},\gamma_{n+1})=0
 \end{aligned}
 \label{eq:trapezoidalTimeScheme}
 \right. .
\end{equation}
These equations are solved iteratively using the Jacobian-free Newton–Krylov method~\cite{Knoll_2004} (see \cite{Boedec_JCP_2017} for details).

\section{Numerical setup and validation} \label{Sec:Val}
\subsection{Dimensionless groups}

Let $V$ and $A$ denote the enclosed volume and the surface area of a deformable particle, respectively. The volume remains constant and defines a length scale $R=(3V/4\pi)^{1/3}$. For a lipid vesicle, the surface area of the membrane also remains constant. Then the reduced volume $\nu = 6\sqrt{\pi}VA^{-3/2}$ ($0 < \nu \le 1$) measures the asphericity of the vesicle. In the present work, we consider only laminar viscous flow through a uniform channel of either circular or rectangular (square) in cross-section. Let $R_t$ denote the characteristic dimension of the flow channel (the radius of a cylindrical tube or the half-width of the cross-section of a rectangular channel), then the ratio $\beta=R/R_t$ measures the flow confinement; the particle's motion is more significantly hindered by particle-wall interactions as the confinement increases. The length-to-diameter (width) ratio of the channel is defined by $\zeta=2\ell_x/(2R_t)$, with $2\ell_x$ being the total length of the channel. 

In the absence of any particle or far from the particle, the flow approaches the unperturbed flow $\bm{u}^{\infty}$ in a channel. It is Poiseuille flow with a parabolic velocity profile for a circular tube \par
\begin{equation}
\bm{u}^{\infty}=2U\left(1-\frac{y^2+z^2}{R^2_t}\right)\bm{e}_x,
\label{eq:PoiseuilleFlowInCapillary}
\end{equation}
where $U=Q/(\pi R_t^2)$ is the mean velocity with $2U$ representing the maximum undisturbed velocity at the centerline of the tube. The unperturbed flow in a rectangular channel is given in Ref.~\cite{Yih_1979}, see also Ref.~\cite{Kuriakose_2011}, that is  \par
\begin{equation}
 \frac{\bm{u}^{\infty}\cdotp\bm{e}_x}{\Upsilon}=\left(\ell^2_z-z^2\right)+\sum^{\infty}_{m=1}B_m\cosh\left(\frac{b_my}{\ell_z}\right)\cos\left(\frac{b_mz}{\ell_z}\right),
 \label{eq:velocityInRectangularChannel}
\end{equation}
with
\begin{equation}
 \Upsilon=-\frac{1}{2\eta^\mathrm{e}}\frac{\mathrm{d}p}{\mathrm{d}x},\quad b_m=\frac{(2m-1)\pi}{2},\quad  B_m=\frac{(-1)^m4~\ell^2_z}{b^3_m\cosh\left(b_m\ell_y/\ell_z\right)},
\end{equation}
where $\ell_y$ and $\ell_z$ denote the channel's half-height and half-width, respectively. 
Integrating over the channel's cross-section yields the volumetric flow rate $Q$, \par
\begin{equation}
 \frac{Q}{\Upsilon}=\frac{8\ell_y\ell^3_z}{3}+\sum^{\infty}_{m=1}B_m\left(\frac{2\ell_z}{b_m}\right)^2\sinh\left(\frac{b_m\ell_y}{\ell_z}\right)\sin\left(b_m\right).
 \label{eq:fluxOverRectangularChannel}
\end{equation}
The mean velocity is $U=Q/(\ell_y\ell_z)$. Substituting Eq.~(\ref{eq:fluxOverRectangularChannel}) into (\ref{eq:velocityInRectangularChannel}) leads to a velocity profile $\bm{u}^{\infty}$ that is proportional to the mean velocity $U$ and depends on the aspect ratio of the channel's cross-section $\ell_y/\ell_z$. In our simulations, we set $m=40$, as in~\cite{Kuriakose_2011}.
The maximum undisturbed velocity at the centerline of a square channel is approximately $2.1U$.

In addition to the dimensionless geometrical parameters mentioned above ($\nu$ and $\beta$), the interfacial mechanical property of a deformable particle immersed in a viscous flow introduces a dynamic dimensionless parameter, the capillary number $\mathit{Ca}=\tau_{\mathit{sr}}/\tau_{\mathit{f}}$, which is the ratio of the characteristic shape relaxation timescale $\tau_{\mathit{sr}}$ to a viscous timescale $\tau_{\mathit{f}}$. Specifically, the capillary number for 
\begin{itemize}
\item surface tension dominant liquid drops, $\mathit{Ca}=\eta^\mathrm{e} U /\gamma$, measures the relative importance of viscous forces to interfacial tension forces;
\item bending dominant membrane of vesicles, $\mathit{Ca}=\eta^\mathrm{e} U \mathit{R}^2/\kappa$, is a ratio of viscous stress to resistive bending stress on the membrane;
\item shearing dominant membrane of capsules, $\mathit{Ca}=\eta^\mathrm{e}U/\mu_s$, determines the relative importance of viscous forces to resistive elastic forces on the membrane. 
\end{itemize}
In this study, we assume that fluid flows at an imposed, constant volumetric flow rate $Q$ driven by a pressure difference between the channel's inlet and outlet. Hence, the dynamical behavior of a deformable particle flowing in a tube or square channel is determined only by four independent dimensionless groups: the confinement $\beta$, the reduced volume $\nu$ (only for vesicles), the capillary number $Ca$, and the viscosity ratio $\lambda$ (which is set to unity, unless specified otherwise).  Numerical solutions should be independent of the total length of the channel provided that it is sufficiently long for the disturbances to become negligibly small at the channel ends. So, we present in the next subsection two numerical examples to show the effects of the channel's length and the minimum element size ($\ell_{\min}$) of the channel mesh, and then give a general criterion for the choice of these two parameters.

\subsection{Effect of the channel's length and wall mesh}

\begin{figure}[!htbp]
 \centering
   \includegraphics[width=1.0\linewidth]{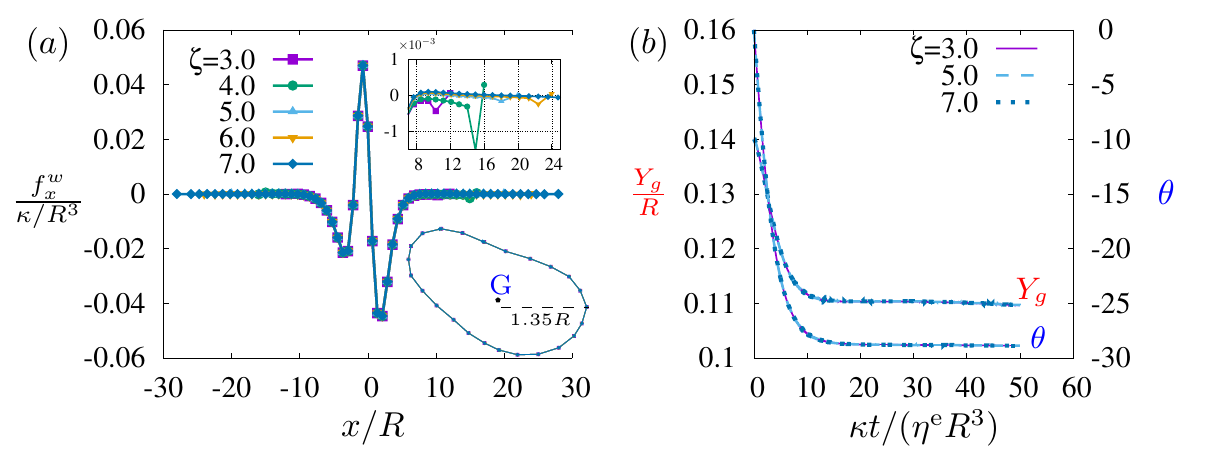} 
 \caption{Effect of the channel length quantified by $\zeta$ for a vesicle in a tube flow with $\ell_{\min}= 0.18 R_t$. The parameters are $\nu=0.9$, $\beta=0.25$, and $Ca=1$.
 (a) The disturbed shear stress on the tube wall $f^w_x$ for different values of $\zeta$. The upper inset is a zoomed-in view. The bottom inset shows the steady vesicle profile in the $xy$-plane with G being the vesicle's centroid.
 (b) Temporal evolution of the centroid ($Y_g/R$) and inclination angle ($\theta$, in degree).}
 \label{fig:TubeMeshLength_Vesicle}
 \end{figure}
 Figure~\ref{fig:TubeMeshLength_Vesicle} shows the numerical results of an initial prolate vesicle ($\nu=0.9$, the membrane surface discretized with $N=320$ Loop elements) flowing in a cylindrical tube with different lengths.
Fig.~\ref{fig:TubeMeshLength_Vesicle}(a) shows the disturbed shear stress $f^w_x$ along the tube wall in the $xy$-plane (see Fig.~\ref{fig:schematicFlow}) for different tube lengths $2\ell_x$ by varying the ratio $\zeta$. As can be seen from this figure, the disturbed shear stress $f^w_x$ decreases exponentially with distance from the vesicle and are vanishingly small  towards the channel ends when $\zeta\geq 5$. This observation is consistent with the analysis of Liron and Shahar~\cite{Liron_1978}, who showed the perturbation flow in tube generated by a point-force distribution decays exponentially with distance from the source point.
Interestingly, the steady-state shape [Fig.~\ref{fig:TubeMeshLength_Vesicle}(a)], as well as the temporal evolution curve of the centroid $Y_g$ and the inclination angle $\theta$ [Fig.~\ref{fig:TubeMeshLength_Vesicle}(b), the angle between the vesicle major axis and the flow direction $\bm{e}_x$], is insensitive to far-field perturbations, which is important when studying the dynamical behavior of a deformable particle in a channel flow. Indeed, the numerical results are virtually indistinguishable when $\zeta\geq 3$.

 \begin{figure}[!htbp]
\centering\includegraphics[width=0.6\linewidth]{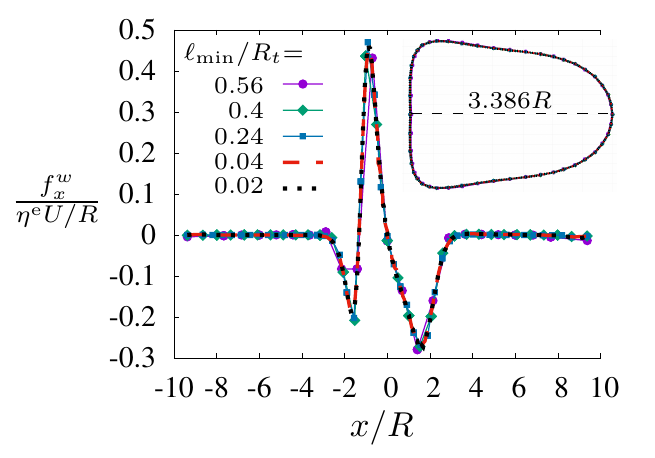}
\caption{The disterbuted shear stress on the tube wall $f^w_x$ for a liquid drop ($N=1280$) for different ratios $\ell_{\min}/R_t$. The parameters are $Ca=0.5$, $\beta=0.8$, and $\zeta=7.5$.
The inset shows the steady drop profile in the $xy$-plane.}
\label{fig:TubeMeshSize_Drop}
\end{figure}

Regarding the minimum element size $\ell_{\min}$ that may be required to obtain accurate results, we have set, in this example, $\ell_{\min}=0.18R_t$, which is close to that used by Hu et al.~\cite{Hu_JFM_2012} in highly confined conditions. Moreover, our test of a liquid drop in a cylindrical tube with a wide range of $\ell_{\min}\in \left[0.02R_t, 0.56R_t\right]$ shows that the resulting $f^w_x$ matches very well at the same grid points in the wall mesh. Again the drop shape remains essentially unaffected, as shown in Fig.~\ref{fig:TubeMeshSize_Drop}.
Even in the coarsest mesh tested, i.e., $\ell_{\min}=0.56R_t$, in which the peak of the wall shear stress is not captured due to a lack of grid points there, the obtained values at the existing grid points are very close to those obtained by finer meshes.  In this study, the general rule is that we set the ratio $\zeta \approx 5$--7 and the dimensionless minimum element size in the wall mesh $\ell_{\min}/R=\min \left[O(h/R), O(0.2\beta^{-1})\right]$, where $h/R$ is the dimensionless gap size between the particle surface and the channel wall. 
 
 \subsection{Numerical validation}

\begin{figure}[!htbp]
 \centering
 \includegraphics[width=1.0\linewidth]{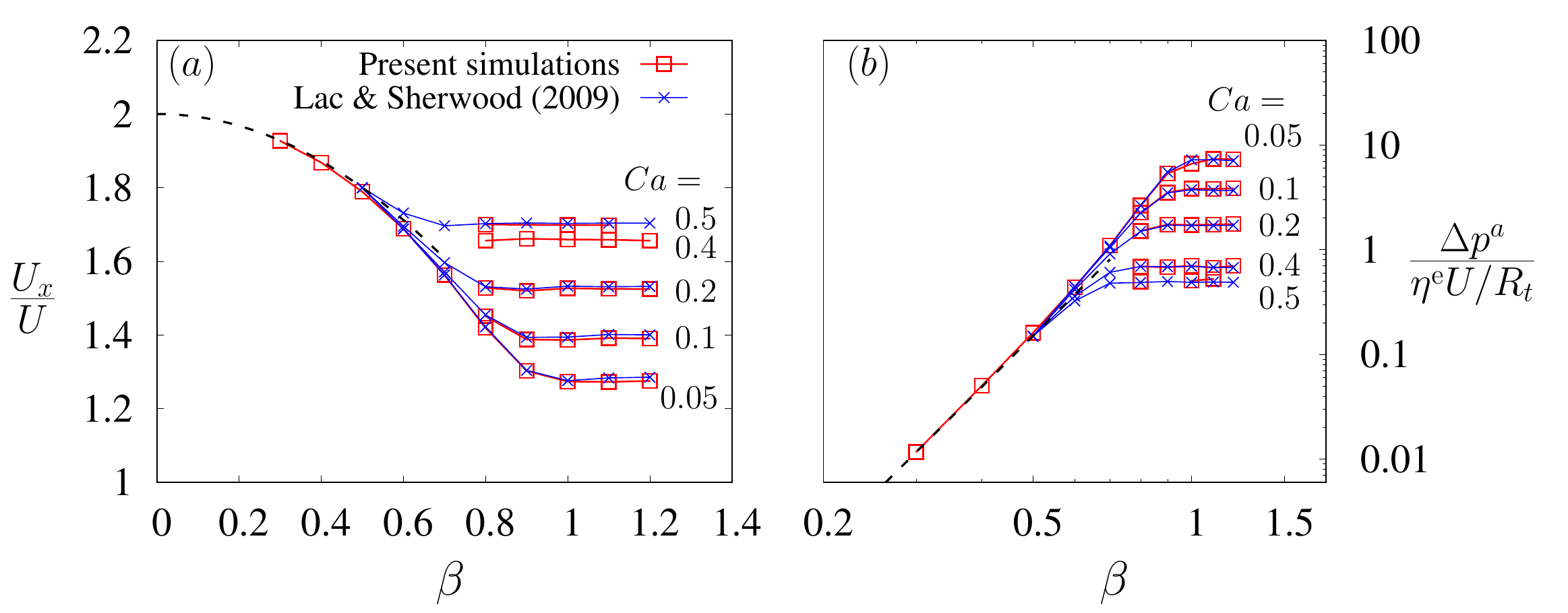}
 \caption{(a) droplet relative velocity $U_x/U$ and (b) dimensionless additional pressure drop $\Delta p^{a}/(\eta^\mathrm{e} U/R_t)$ as a function of the confinement $\beta$ compared to those reported in Ref.~\cite{Lac_JFM2009}. The dashed curves are the  theoretical predictions of \eqref{eq:dropInTube_velocityLowConfinement} and~\eqref{eq:dropInTube_pressureLowConfinement} for $\beta \ll 1$.}
 \label{fig:Lac2009_dropInTube}
\end{figure}

We validate the coupled isogeometric FEM-BEM approach by comparing the simulation results with a well-know example of a (clean, surfactant-free) liquid drop in tube flow, for which very highly accurate numerical computations are available in the literature (e.g., Ref.~\cite{Lac_JFM2009}). The motion of the drop for given confinement $\beta$ is determined only by the capillary number $Ca$ (apart from the viscosity ratio $\lambda$, which is set to unity for this comparison). Of particular interest are the drop relative velocity $U_x/U$ and the dimensionless additional pressure drop $\Delta p^{a}/(\eta^\mathrm{e} U/R_t)$, the latter is due to the presence of the drop in tube flow in order to maintain the volumetric flow rate $Q=\pi R_t^2 U$. Our 3D numerical results (with $N=320$ elements) are compared with the axisymmetric simulations reported in~\cite{Lac_JFM2009}. The comparison in Fig.~\ref{fig:Lac2009_dropInTube} shows excellent agreement. Under weak confinement (i.e., small $\beta$), the simulation results are also in excellent agreement with the theoretical predictions for a vanishingly small droplet moving along the centerline of a tube~\cite{Brenner_1970,Hetsroni_JFM_1970}, given by \par

\begin{subequations} \label{E:smalltube}
  \begin{gather}
   \frac{U_x}{U}=2-\frac{4}{5}\beta^2 + O(\beta^3),
 \label{eq:dropInTube_velocityLowConfinement} \\
\frac{R_t}{\eta^\mathrm{e} U}\Delta p^a =\frac{24}{5}\beta^5 + O(\beta^{10}).
 \label{eq:dropInTube_pressureLowConfinement}
 \end{gather}
\end{subequations}

\begin{figure}[!htbp]
 \centering\includegraphics[width=1.0\linewidth]{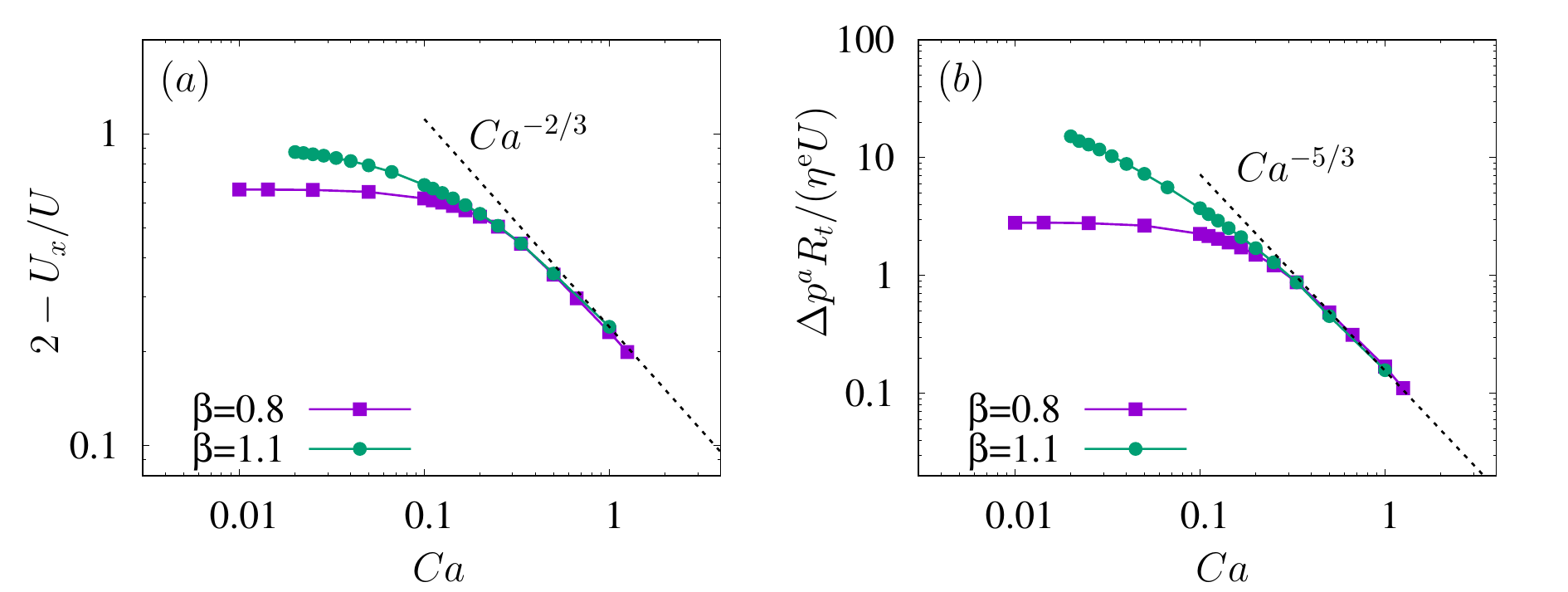}
 \caption{(a) droplet relative velocity $2-U_x/U$ and (b) dimensionless additional pressure drop $\Delta p^{a}/(\eta^\mathrm{e} U/R_t)$ as a function of the capillary number $Ca$ for $\beta=0.8$ and 1.1. The dashed curves are the scalings of \eqref{eq:dropInTube_velocityHighCa} and~\eqref{eq:dropInTube_pressureHighCa} at high $Ca$.}
 \label{fig:UmigDelP_dropInTube}
 \end{figure}

As an additional verification and validation of the numerical model, we plot, in Fig.~\ref{fig:UmigDelP_dropInTube}, the effect of the capillary number $Ca$ on the drop velocity and the additional pressure drop for two values of confinement $\beta=0.8$ and 1.1. Fig.~\ref{fig:UmigDelP_dropInTube}(a) shows how much the drop velocity exceeds the mean velocity of the suspending fluid, approaching the axis velocity $2U$ and resulting in a dramatic decrease in the additional pressure drop [Fig.~\ref{fig:UmigDelP_dropInTube}(b)] as $Ca$ increases. At high capillary numbers, the dimensionless groups $2-U_x/U$ and $\Delta p^{a}/(\eta^\mathrm{e} U/R_t)$ exhibit a remarkable power law. Lac and Sherwood~\cite{Lac_JFM2009} provided via the asymptotic analysis for a long slender drop in tube flow the following scalings at high $Ca$ \par
\begin{subequations} \label{E:highCa}
  \begin{gather}
   2-U_x/U \sim Ca^{-2/3},  \label{eq:dropInTube_velocityHighCa} \\
\frac{R_t}{\eta^\mathrm{e} U}\Delta p^a \sim Ca^{-5/3}.
 \label{eq:dropInTube_pressureHighCa}
 \end{gather}
\end{subequations}
As shown, the present 3D simulations captured these limiting behaviors.

\begin{figure}[!htbp]
 %\centering\includegraphics[width=\linewidth]{../fig/numMethod/validation/dropSpaceGeo.pdf}
\includegraphics[width=0.9\linewidth]{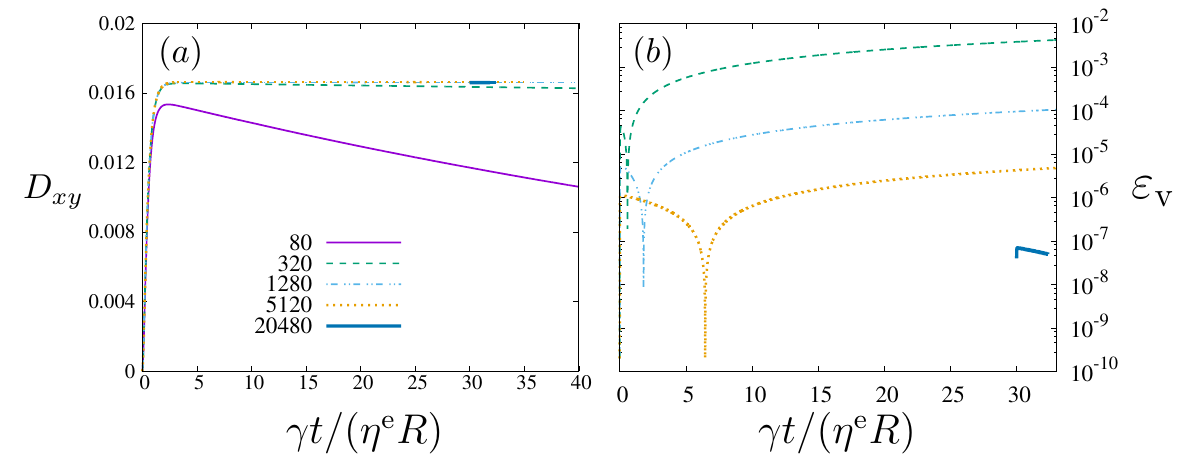}
 \caption{Temporal evolution of (a) the Taylor deformation parameter $D_{xy}$ and (b) the relative derivation  of the enclosed volume $\varepsilon_{\mathrm{v}}=V/V_0-1$ of a drop in a 
 capillary flow ($Ca=0.05$ and $\beta=0.8$) as a function of the number of elements $N$ used on the drop surface.}
 \label{fig:dropSpaceGeo}
\end{figure}

We then proceed to conduct a convergence study on the spatial and  temporal discretization for one of the above settings at $Ca=0.05$ and $\beta=0.8$. The deformation behavior of the drop is characterized by the Taylor deformation parameter $D_{xy}=(L-B)/(L+B)$, where $L$ and $B$ are the major and minor axis of the drop profile in the $xy$-plane. The temporal evolution of $D_{xy}$ plotted in Fig.~\ref{fig:dropSpaceGeo}(a) shows that except for the coarsest mesh (80 elements), which gives a non-converged solution, the simulations from other numbers of mesh elements ranging from 320 to 20480 lead to very good agreement results. We assess the effect of mesh refinement by examining the enclosed volume $V$ and its drift from the initial volume $V_0$: $\varepsilon_{\mathrm{v}}=(V-V_0)/V_0$. It is seen that the volume drift is only 0.4\% after a long-time simulation (i.e., $\gamma t/(\eta^\mathrm{e} R)=30$) with 320 elements. Its temporal evolution (for $N$ ranging from 320 to 20480) displayed in Fig.~\ref{fig:dropSpaceGeo}(b) suggests that one more subdivision process (i.e., multiplying the number of elements by four) leads to at least one order of magnitude gaining in volume conservation.  We performed these simulations from an initially spherical drop using the RKF45 time-stepping scheme (i.e., adaptive time step scheme), while the computations with the finest mesh (20480 elements) started from a drop shape obtained with 5120 elements. 

\begin{figure}[!htbp]
 %\centering\includegraphics[width=0.6\linewidth]{../fig/numMethod/validation/dropSpaceConvergTime_D.pdf}
  \centering\includegraphics[width=0.6\linewidth]{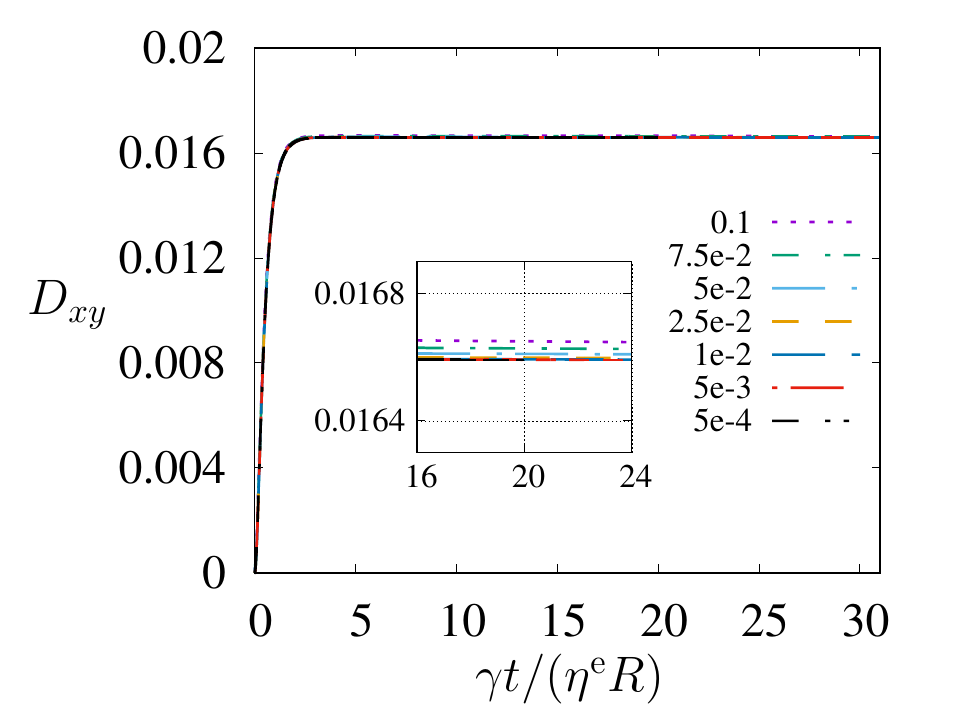}
 \caption{Temporal evolution of the Taylor deformation parameter $D_{xy}$ of a drop ($N=1280$ elements) flowing in a capillary flow ($\mathrm{Ca}=0.05,~\beta=0.8$) as a function of time steps (scaled by $\eta^{\mathrm{e}} R/\gamma$). Time-stepping scheme used is the trapezoidal rule with a fixed time step.}
 \label{fig:dropSpaceConvergTime_D}
\end{figure}

Using $N=1280$ elements, we assess the effect of temporal discretization by varying the time step from 0.1 to $5\times 10^{-4}$. As can be seen from Fig.~\ref{fig:dropSpaceConvergTime_D}, the implicit time integration with these time steps leads to a consistent, very good agreement result. 
 
\begin{figure}[!htbp]
 %\centering\includegraphics[width=\linewidth]{../fig/numMethod/validation/convergence.pdf}
 \centering\includegraphics[width=\linewidth]{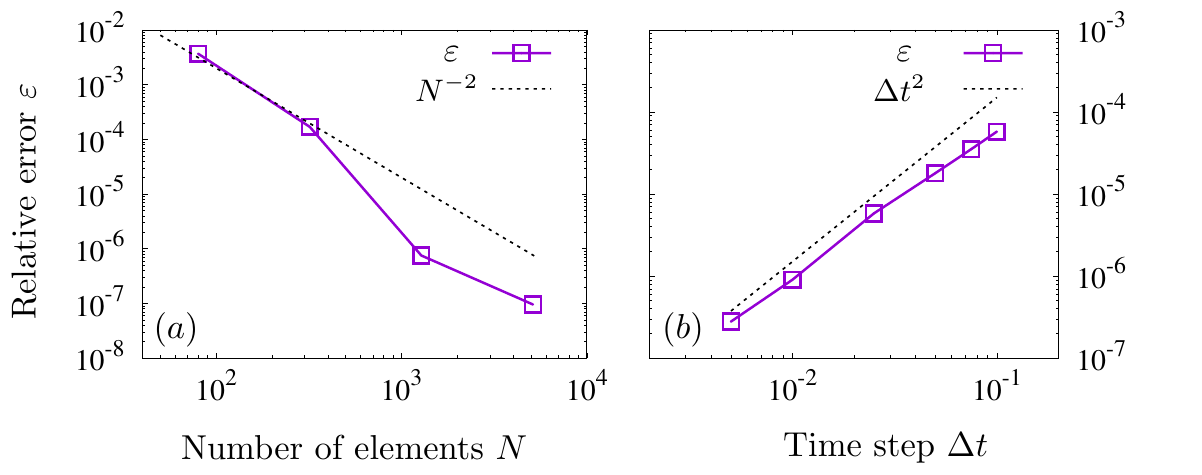}
 \caption{Relative error of the Taylor deformation parameter of a drop in a capillary flow ($Ca=0.05$ and $\beta=0.8$) as a function of (a) the number of elements $N$ and (b) the time step size $\Delta t$ (scaled by $\eta^{\mathrm{e}} R/\gamma$).}
 \label{fig:convergence}
\end{figure}

To conclude this validation subsection, we provide an estimate of the convergence rate of the numerical method. The convergence order of spatial and temporal discretization is evaluated via the relative error of the Taylor deformation parameter $\varepsilon=(D_{xy}-D_\mathrm{ref})/D_\mathrm{ref}$ at a dimensionless time $\gamma t/(\eta^{\mathrm{e}} R)=30$ (a steady state), where $D_\mathrm{ref}$ is the reference Taylor deformation parameter. Fig.~\ref{fig:convergence} makes it clear that the present algorithm preserves second-order convergence in both space and time for a liquid drop confined in capillary flow.  For the spatial convergence shown in Fig.~\ref{fig:convergence}(a), the RKF45 scheme is used, and the reference value is computed with 20480 elements. For the temporal convergence displayed in Fig.~\ref{fig:convergence}(b), the trapezoidal scheme is used with 1280 elements, and the reference value is computed with a time step $\Delta t=5\times 10^{-4}$ (scaled by $\eta^{\mathrm{e}} R/\gamma$). The previous study~\cite{Boedec_JCP_2017} on unbounded soft particles shows that the second-order convergence is not affected by the membrane's constitutive law, including bending stiffness, we may expect that the second-order convergence achieved for confined liquid drops in tube flow is also applicable to other confined soft particles.

\section{Numerical examples}\label{Sec:Exp}
In this section, we present further simulation results to demonstrate the accuracy and stability of the numerical method. Having described drop dynamics in Section~\ref{Sec:Val}, we focus now on the other three types of soft particles to illustrate potential applications. Specifically, we simulate (i) an elastic capsule in a square channel, (ii) a vesicle in a cylindrical tube, and (iii) a single RBC in a capillary.
Where it is possible, we compare the numerical results with previously published studies.

\subsection[Capsule in square channel]{Elastic capsule in a square channel}
\label{sec:capsuleInSquareChannel}
\begin{figure}[!htbp]
 %\centering\includegraphics[width=0.7\linewidth]{../fig/capsule/flowSchema.pdf}
  \centering\includegraphics[width=0.7\linewidth]{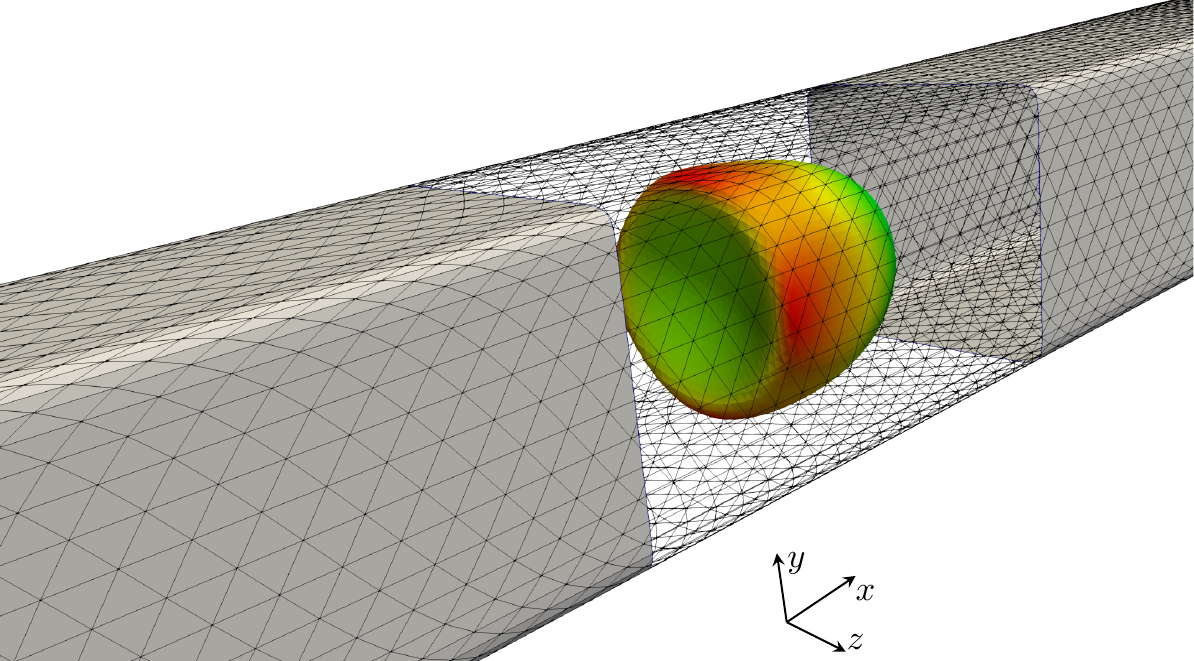}
 \caption{An elastic capsule flowing in a square microchannel. The channel wall mesh is generated by Loop's subdivision process and rounded with an arc-circle. Colors on the surface of the capsule represent the $x$-component of the membrane elastic force ($f^m_x$).}
 \label{fig:capsuleInTube}
\end{figure}

As shown in  Fig.~\ref{fig:capsuleInTube}, the first simulation example concerns the steady-state deformation of an initially spherical capsule moving through a square microchannel with the undisturbed flow in the channel $\bm{u}^{\infty}$ being given by~\eqref{eq:velocityInRectangularChannel}.

\begin{figure}[!htbp]
 %\centering\includegraphics[width=0.8\linewidth]{../fig/capsule/capsuleInTube_differntCa.pdf}
  \centering\includegraphics[width=0.8\linewidth]{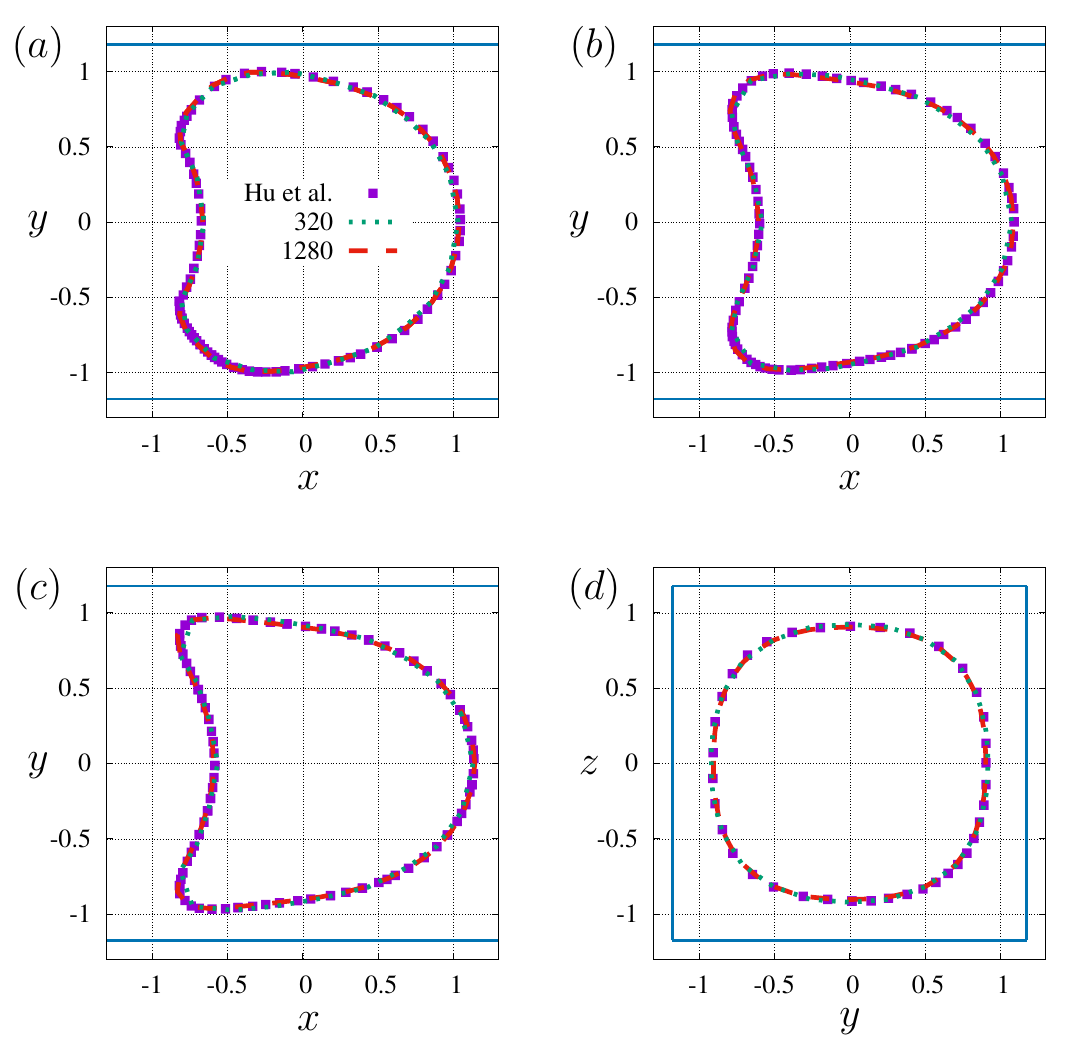}
 \caption{Comparison of the steady-state deformation of a capsule flowing in a square channel with those obtained by Hu et al.~\cite{Hu_2013} for $\beta=0.85$. Profiles in the $xy$-plane: (a) $Ca=0.02$, (b) $Ca=0.05$, and (c) $Ca=0.1$. Profile in the $yz$-plane: (d) $Ca=0.1$. The blue lines indicate channel walls.}
 \label{fig:capsuleInTube_differntCa}
\end{figure}

We performed computations with the strain-hardening Sk law \eqref{eq:energyDensityForCapsule} with $C=1$, as used in~\cite{Hu_2013},  for $Ca=$ 0.02, 0.05, and 0.1, and at $\beta=0.85$. Simulations are run for two capsule meshes consisting of 320 and 1280 Loop elements. The capsule profiles at steady-state  are displayed in Fig.~\ref{fig:capsuleInTube_differntCa}(a)-(c) in the $xy$-plane  and in Fig.~\ref{fig:capsuleInTube_differntCa}(d) in the $yz$-plane. Increasing the capillary number, or equivalently decreasing the membrane elasticity, leads to the capsule less able to retain its spherical shape and more elongated. The obtained results are compared with those reported in Ref.~\cite{Hu_2013} in which the surface of the capsule is discretized by 1280 quadratic triangular element~\cite{Ramanujan_1998}.  Thanks to the use of Loop elements for the representation of geometry and the two solvers in our simulations, even a coarse mesh of 320 elements  reproduces the numerical results of \cite{Hu_2013}, which represents a significant improvement.

\subsection{Vesicle in a circular tube}

\begin{figure}[!htbp]
% \centering\includegraphics[trim=10 100 10 10,clip,width=0.7\linewidth]{../fig/vesicle/schematic_vesicleInTube.png}   
 \centering\includegraphics[width=0.7\linewidth]{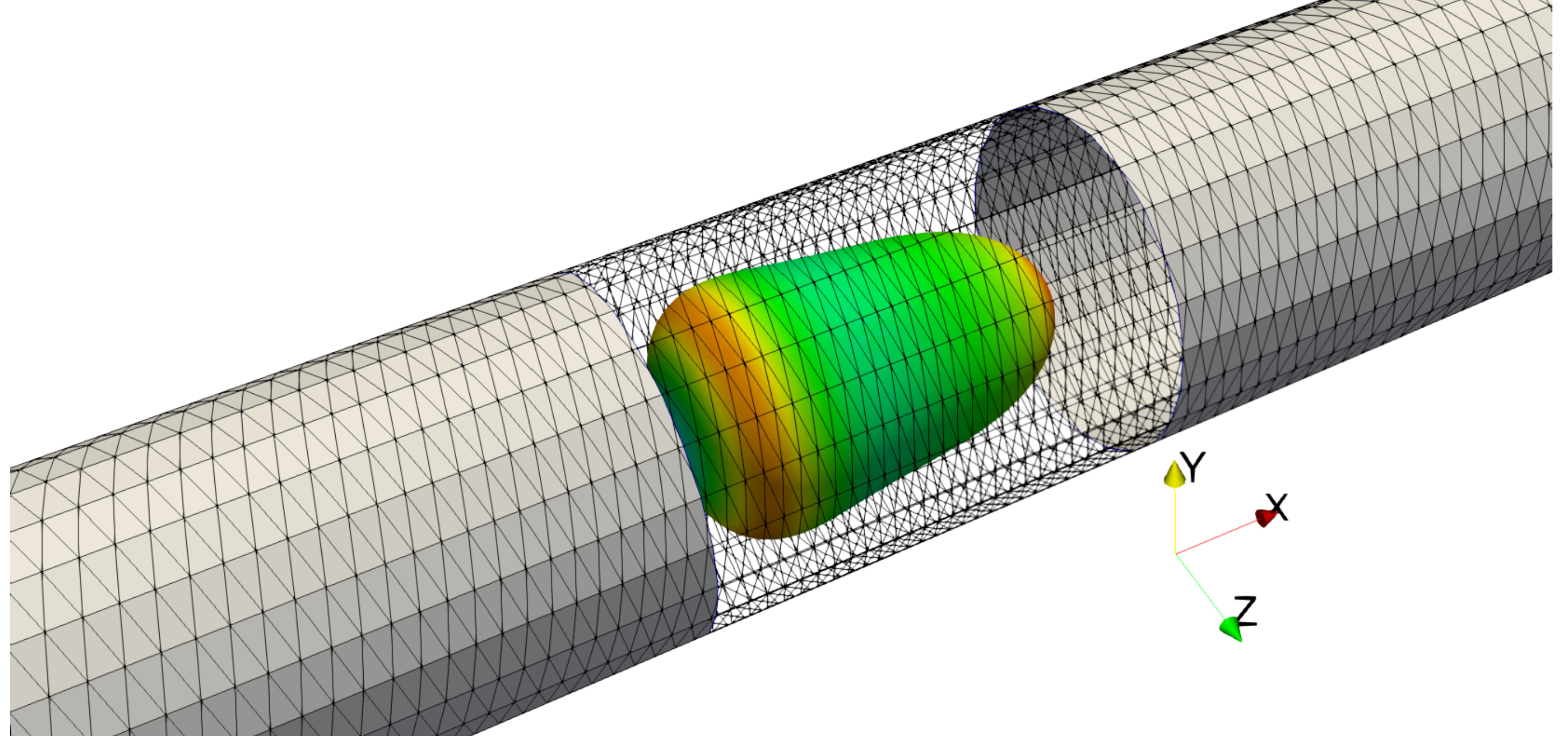}
 \caption{Deformation of a vesicle in tube flow for $Ca=10$, $\nu=0.9$, and $1/\beta=1.2$. Colors on the vesicle surface represent the local mean curvature.}
 \label{fig:schematic_vesicleInTube}
\end{figure}

The second numerical example deals with a confined vesicle flowing through a circular tube, as illustrated in Fig.~\ref{fig:schematic_vesicleInTube}.
In aqueous solution, lipid vesicles exhibit a large variety of shapes and shape transformations, in particular, they can exhibit a biconcave shape typical of red blood cells. When confined in capillary tubes subjected to Poiseuille flow, however, vesicles assume complex shapes and behave in different ways due to an intricate interplay between flow stresses, membrane's bending rigidity, and confinement~\cite{Farutin_2014,Chen_2019}.

\begin{figure}[!htbp]
 %\centering\includegraphics[width=0.7\linewidth]{../fig/vesicle/effectBeta_Ca10nu0d9.pdf}
  \centering\includegraphics[width=0.7\linewidth]{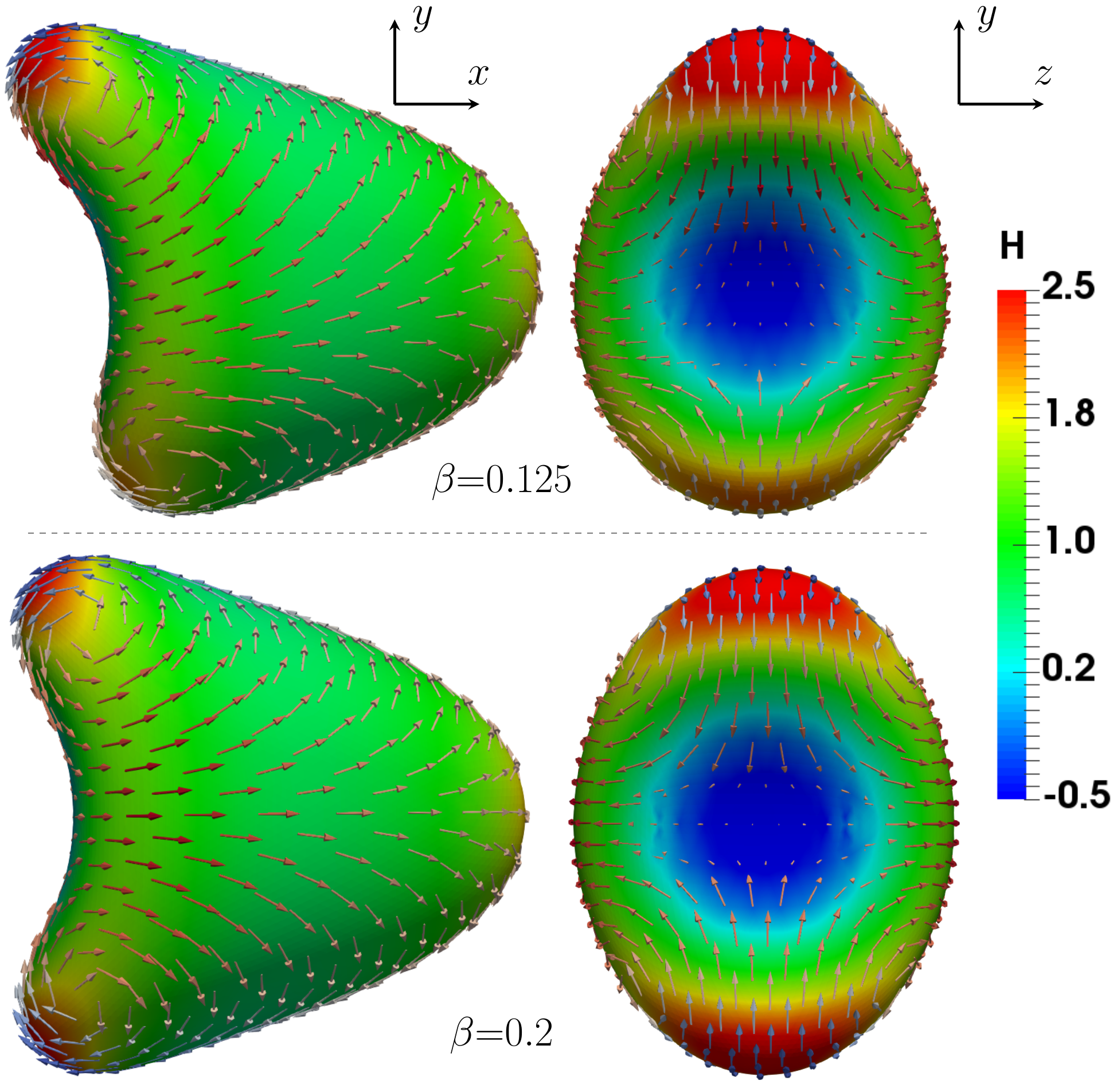}
 \caption{The steady-state slipper shape (top, $\beta=0.125$) and croissant shape (bottom, $\beta=0.2$) of a vesicle ($\nu=0.9$, discretized by 1280 elements)  in a confined axial Poiseuille flow ($Ca=10$). Left: side view in the $xy$-plane; right: rear view in the $yz$-plane. Colors on the vesicle surface represent the local mean curvature, and the arrows show the membrane flow.}
 \label{fig:effectBeta_Ca10nu0d9_vesicle}
\end{figure}

To illustrate two types of 3D vesicle shapes at steady state in a confined axial Poiseuille flow, we performed computations for an initially prolate vesicle of the reduced volume $\nu=0.9$ at two confinement conditions, i.e., $\beta=0.125$ and 0.2. The vesicle, initially located at a height $H_0$ ($=0.06$, refer to Fig.~\ref{fig:RBC_iniConfig}) from the flow axis, rapidly changes its shape, becoming a slipper shape, which is characterized by a single mirror symmetry in the $yz$-plane due to the flow curvature. At very weak confinement (i.e., $\beta=0.125$), the slipper shape reaches a steady-state with the inward migration ending at a certain position to the centerline $H=\sqrt{y^2+z^2}\approx 0.028$.  This final stationary shape, as well as its membrane flow structure, is illustrated in the top panel of Fig.~\ref{fig:effectBeta_Ca10nu0d9_vesicle}. The membrane flow is characterized by two unequal vortices both on the front and rear faces of the membrane. At slightly high confinement ($\beta=0.2$), the transitional slipper shape is unstable, becoming a croissant shape characterized by two mirror symmetries in the $yz$-planes, and its radial position $H$ decrease to zero. The membrane flow now consists of the two equal vortices both on the front and rear faces of the membrane, as illustrated in the bottom panel of Fig.~\ref{fig:effectBeta_Ca10nu0d9_vesicle}.
These are long-time simulations  and steady state using the implicit time integration is reached around  $\kappa t/(\eta^eR^3)\approx 300$ with a dimensionless time step $\Delta t=4 \times 10^{-3}$. The relative error in the enclosed volume and the total surface area are respectively $\varepsilon_{\mathrm{v}} \approx 0.1\%$  and  $\varepsilon_{\mathrm{A}} \approx 0.045\%$ (for $\beta=0.125$), and $0.07\%$ (for $\beta=0.2$), indicating the high accuracy and stability of the present algorithm.

 \begin{figure}[!htbp]
  \centering\includegraphics[width=0.9\linewidth]{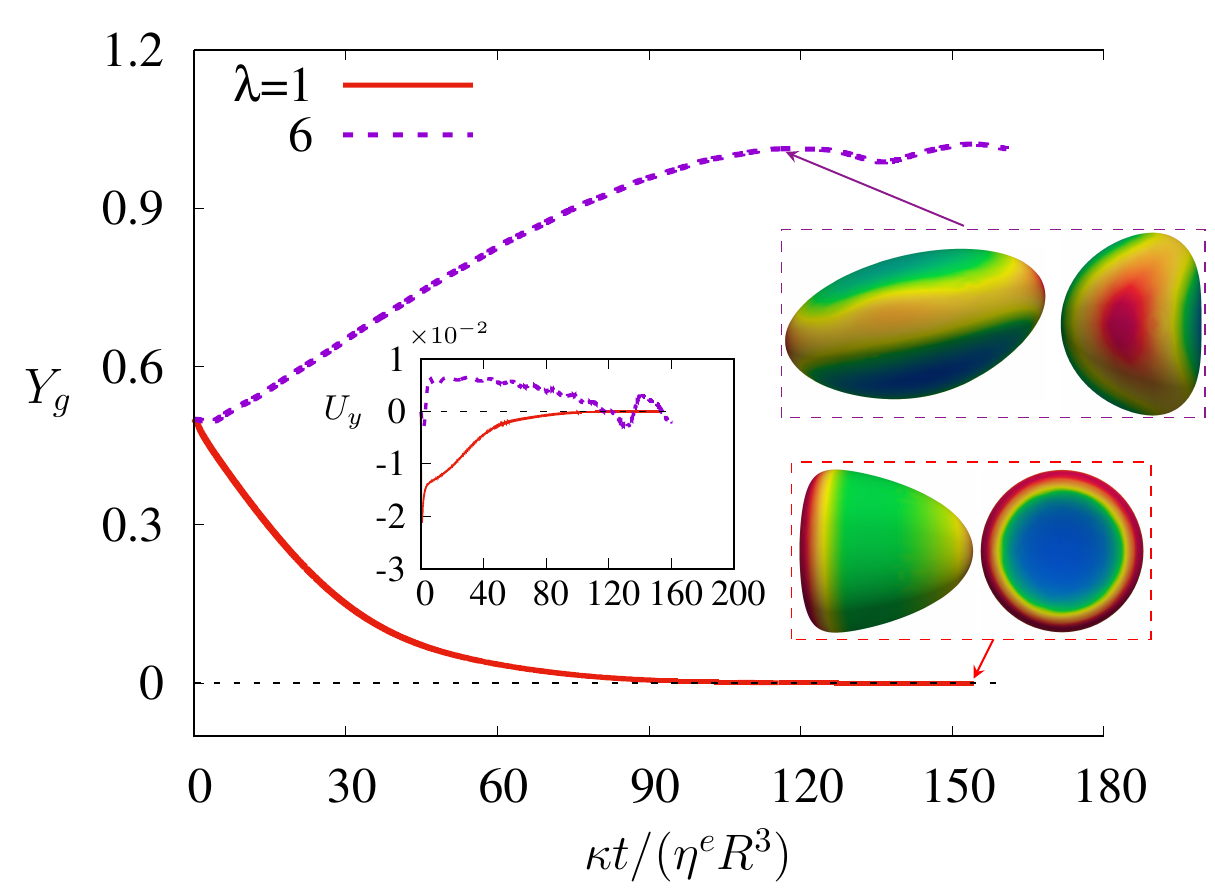}
 \caption{Temporal evolution of the centroid $Y_g$ of a vesicle ($\nu=0.9$) in a weakly confined ($\beta=0.1$) Poiseuille flow ($Ca=100$) for two different viscosity ratios: $\lambda=1$ and $\lambda=6$. The inset shows the evolution of the lateral migration velocity $U_y$ (left) and vesicle shapes from side and rear views (right) with colors  representing the local mean curvature.}
 \label{fig:case_viscosityContrats}
\end{figure}

 For the sake of simplicity, all these numerical examples are limited to a viscosity ratio of unity (i.e., $\lambda=1$). To demonstrate the ability of the present code to handle non-unity viscosity ratios, we show in Fig.~\ref{fig:case_viscosityContrats} simulation results for a vesicle in a weakly confined Poiseuille flow at $\lambda=1$ and 6.  In the absence of viscosity contrast, the vesicle migrates towards the center (i.e., $Y_g\rightarrow 0$ and $U_y\rightarrow 0$), and an axisymmetric parachute shape is obtained, as illustrated by the insets surrounded by red dashed lines. While the situation is different for $\lambda=6$, starting from an initial position $H_0=0.5R$, the vesicle migrates outwards, and an asymmetric shape is now produced, as shown by the insets surrounded by a purple dashed lines. These simulations results are consistent with a previous study~\cite{Farutin_2014}, which has shown that depending on the viscosity contrast $\lambda$, a vesicle  in an unbounded Poiseuille flow can migrate either inward towards the center, or outward of the flow at high $Ca$ if the initial position is chosen sufficiently far  from the centerline. We note that the effect of the viscosity ratio on the dynamics of a confined vesicle is the subject of a very recent study~\cite{agarwal2020stable}.

\subsection{RBC in capillary flow} 
\label{sec:RBCInCapillaryFlow}
\begin{figure}[!htbp]
 \centering\includegraphics[width=0.6\linewidth]{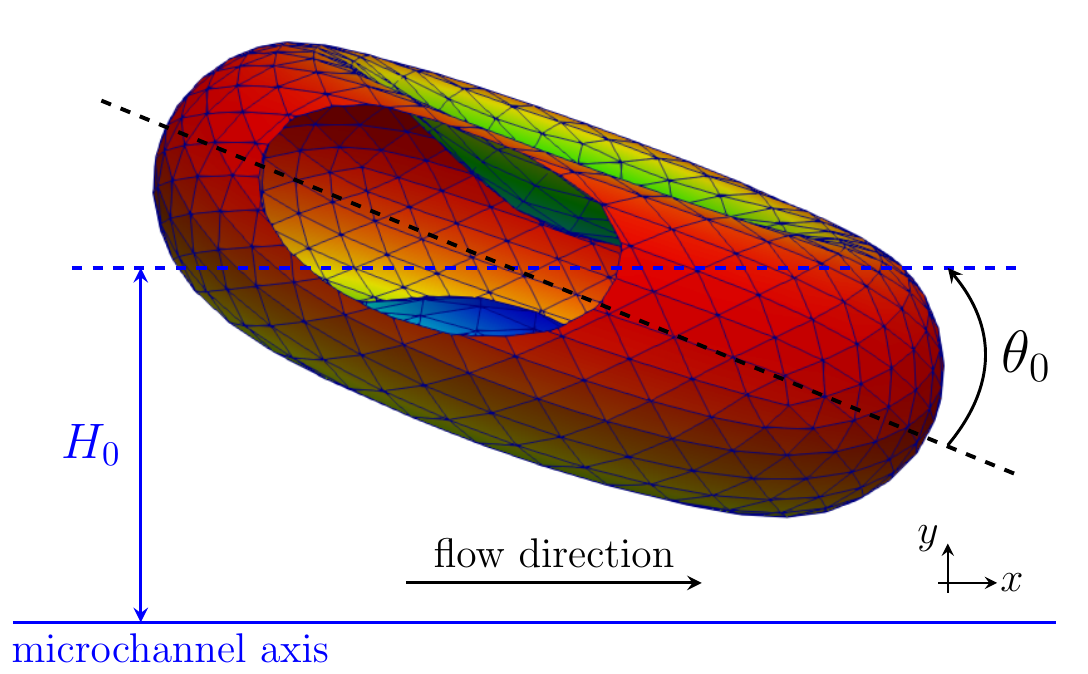}
 \caption{Schematic representation of the initial configuration of a biconcave red blood cell (colors on the membrane represent $|\bm{x}|$), where $H_0$ and $\theta_0$ are respectively the initial offset position of its centroid relative to the flow axis and the initial inclination angle measured from its flat plane to the axis of the flow. }
 \label{fig:RBC_iniConfig}
\end{figure}

The last numerical example concerns a single RBC in capillary flows. The initial biconcave discoid shape of RBC is given by the following expression~\cite{ Evans_1980} \par
\begin{equation}
 {y= \pm  D \sqrt{1 - \frac{4(x^2+z^2)}{D^2}}\left [ a_1 + a_2 \frac{x^2+z^2}{D^2}+ a_3 \frac{\left (x^2+z^2 \right)^2}{D^4}\right ]},
 \label{eq:RBCBiconcaveShape}
\end{equation}
where $D=$~\SI{7.82}{\um} is the cell diameter, $a_1=0.0518$, $a_2=2.0026$ and $a_3=-4.491$. The initial shape is shown in Fig.~\ref{fig:RBC_iniConfig} with a clip to better represent its three-dimensional structure. The volume and surface area of the corresponding RBC are respectively \SI{94}{\micro\metre}$^3$ and \SI{135}{\micro\metre}$^2$, giving a reduced volume $\nu=0.64$ and an effective diameter $D_{\mathrm{eff}}=\sqrt[3]{6V/\pi}=$~\SI{5.64}{\um}. As in the case of a vesicle flowing in capillary, Poiseuille flow is given by Eq.~\eqref{eq:PoiseuilleFlowInCapillary}, where $U=$~\SI{0.016}{\centi\meter\per\second} for all cases considered in this subsection,
which lies in the range of \SIrange{0.001}{1}{\centi\meter\per\second} exploited by Pozrikidis~\cite{Pozrikidis_2005}.
For the membrane with shape memory, the reference shape (i.e., unstressed shape) is another influencing ingredient to compute the elastic force [Eq.~\eqref{eq:edgeSpringForce}]. Here, the initial biconcave form ~\eqref{eq:RBCBiconcaveShape} is used as the unstressed shape.

\begin{figure}[!htbp]
 %\centering\includegraphics[width=0.5\linewidth]{../fig/RBC/caseA/3D_shapes/finalStbale3DShape.png}
  \centering\includegraphics[width=0.7\linewidth]{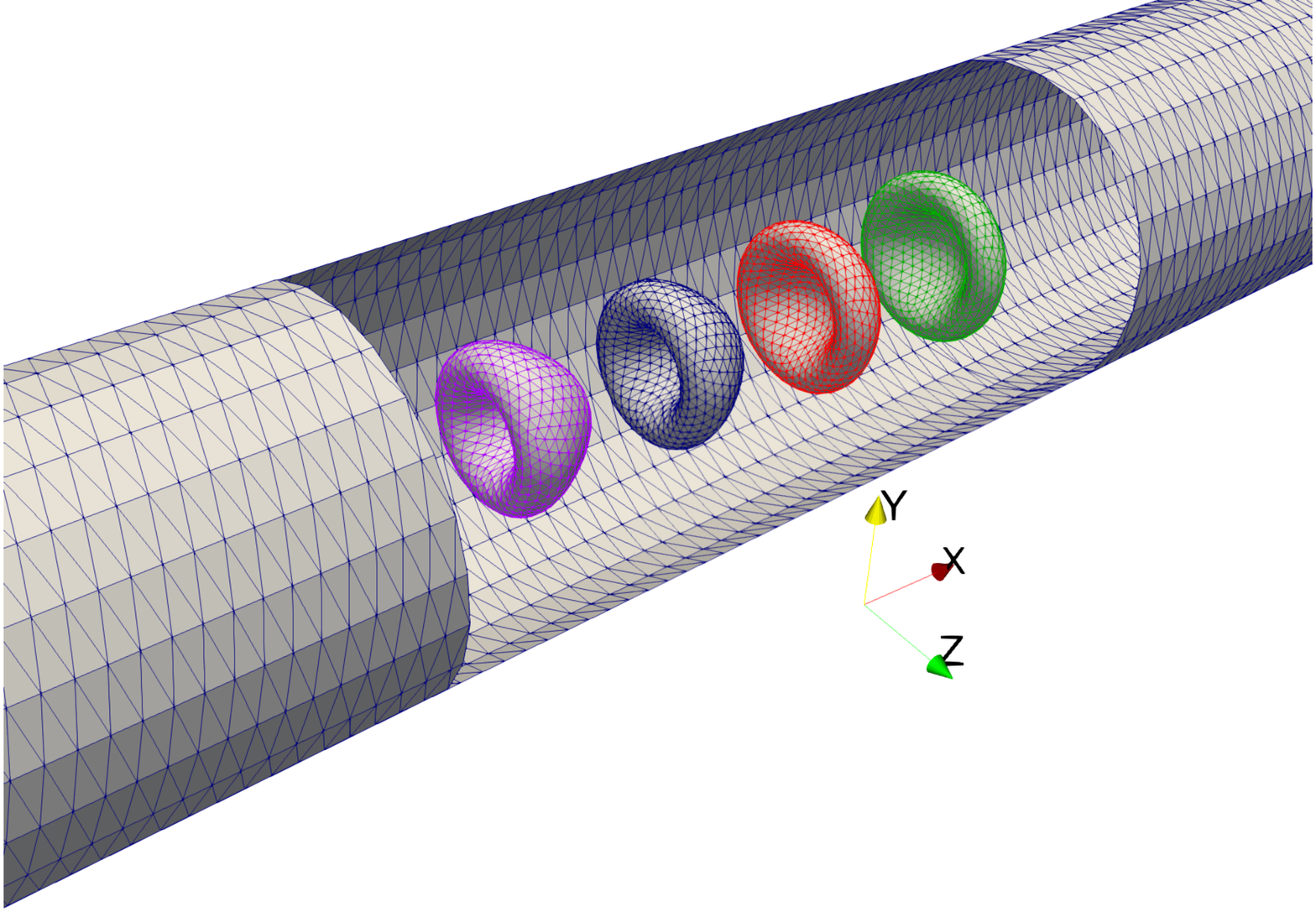}
 \caption{The steady shapes of RBCs in a capillary for four different shear moduli, 0.0, 0.5, 5.0 and 10.0~\si{\micro\newton\per\meter} (from left to right).}
 \label{fig:finalStbale3DShape_RBC_caseA}
\end{figure}

First, we consider that RBCs are initially placed on the axis of the capillary ($H_0=0.0$, $\beta=0.5$) and that their flat surfaces are orthogonal to the flow ($\theta_0=$~\SI{-90}{\degree}) with varying elastic moduli $\mu_s = 0.0, 0.5, 5.0$ and~\SI{10.0}{\micro\newton\per\meter}. The steady shapes, shown in Fig.~\ref{fig:finalStbale3DShape_RBC_caseA}, suggest that the cell deformation is significantly reduced at higher resistance to elastic forces since an RBC with a higher shear modulus has a greater ability to withstand hydrodynamic stresses.

\begin{figure}[!htbp]
% \centering\includegraphics[width=0.8\linewidth]{../fig/RBC/caseB/RBC_shpaes.pdf}
 \centering\includegraphics[width=0.8\linewidth]{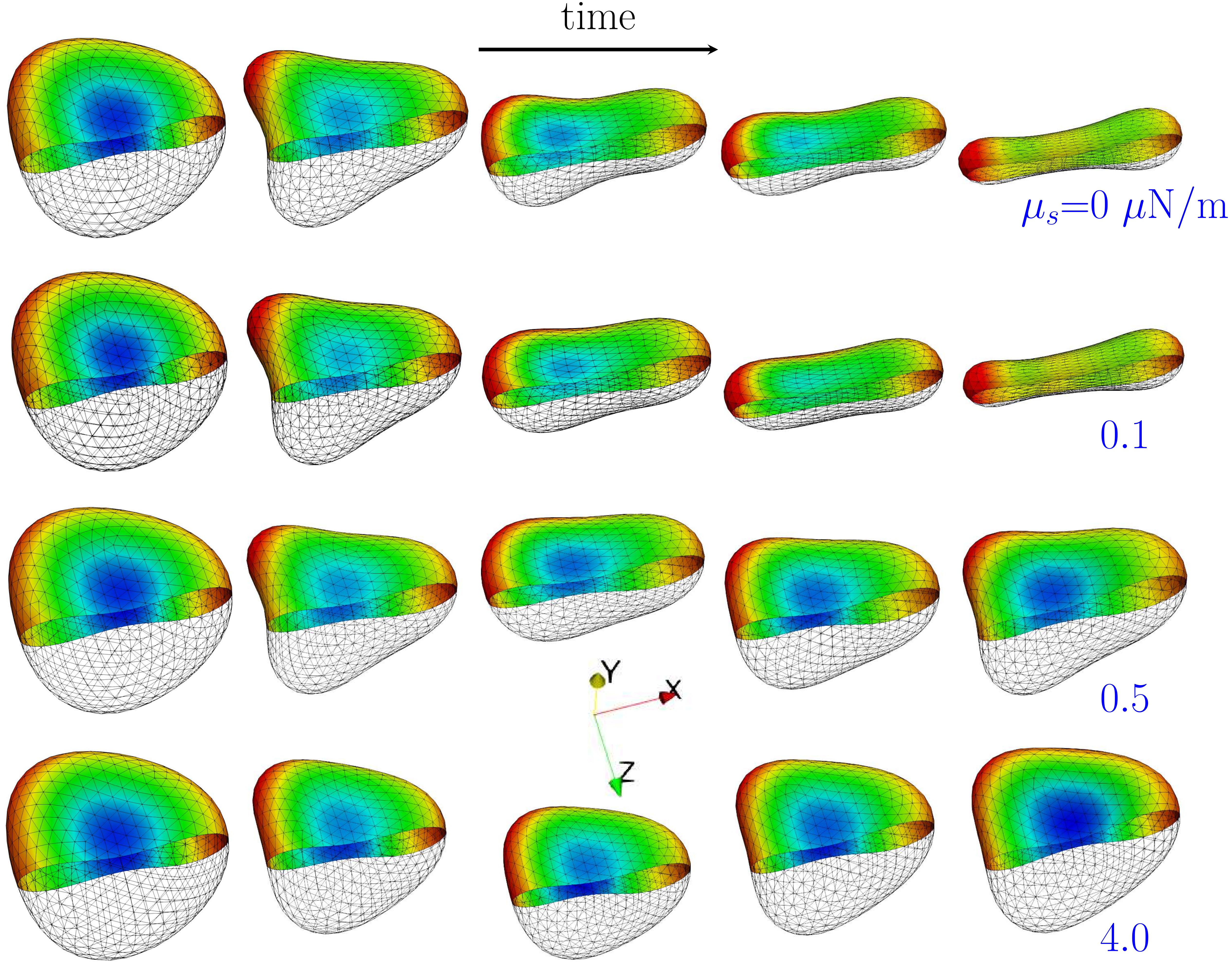}
 \caption{Temporal evolution of the RBCs shape in a capillary flow ($\beta=0.4$) for five different shear moduli $\mu_s=0.0, 0.1, 0.5$, and $4.0$~\si{\micro\newton\per\meter} with $H_0=0.3$ and $\theta_0\approx$~\SI{-6}{\degree}. 
 Membranes are colored by the local mean curvature.}
 \label{fig:RBC_shpaes_in_capillary_H0d3}
\end{figure}

We then consider a case in which the flat plane of RBCs is not placed orthogonally to the flow direction, but only with a small inclined angle $\theta_0\approx$~\SI{-6}{\degree}, and is placed
at  $H_0=0.3$ in a capillary with $\beta=0.4$. By varying the shear modulus of the membrane $\mu_s$ from 0~\si{\micro\newton\per\meter} (vesicle) to a relatively high value 4.0~\si{\micro\newton\per\meter}, 
we obtain a totally different shape evolution process, which depends upon the shear modulus, as shown in Fig.~\ref{fig:RBC_shpaes_in_capillary_H0d3}.

The snapshots in the first row of Fig.~\ref{fig:RBC_shpaes_in_capillary_H0d3} show that an initially biconcave vesicle ($\nu\approx0.64$) placed at $H_0=0.3$ evolves into a slipper shape. The overall evolution of the shape remains essentially unchanged up to $\mu_s \approx 0.1$~\si{\micro\newton\per\meter}. A different transition occurs by increasing the shear modulus to 0.5~\si{\micro\newton\per\meter}, as shown in the third row of Fig.~\ref{fig:RBC_shpaes_in_capillary_H0d3}; a biconcave shape undergoes a transition 
into a biconcave-croissant shape (a biconcave shape with two planes of symmetry like a croissant shape), for which the two dimples are preserved during the transition. The biconcave shape is first stretched under the action of the flow stresses, but the subsequent stretching is mitigated by the cytoskeletal forces, which are characterized by the cytoskeletal shear modulus $\mu_s$.

%Fig.~\ref{fig:RBC_beta0d4} shows the temporal evolution of the lateral position of the centroid $Y_g$ and the inclination angle $\theta$ of RBCs with the shear moduli ranging from 0 (a vesicle) to 4~\si{\micro\newton\per\meter}. These are very long-time simulations as reflected in the plotted dimensionless time scale. It is seen that for the biconcave-croissant shapes (solid lines, $\mu_s \ge 0.5$~\si{\micro\newton\per\meter}), the stable lateral position $Y_g$ increases with the shear modulus $\mu_s$ and the inclination angle $\theta$ has almost zero value (i.e., $\theta \approx$~\SI{1.5}{\degree}). This means that the stable biconcave-croissant shape has its flat plane parallel to the flow direction. While for the two cases with a slipper as final stable shape, the centroid $Y_g$ is still evolving. This case shows how the elasticity of the membrane affects the shape evolution in a confined Poiseuille flow, which may be interesting for further studying the dynamics of capsules in Poiseuille flow~\cite{agarwal2020stable}.

\section{Conclusions}\label{Sec:Concl}
We have presented a coupled isogeometric FEM and BEM computational framework to accurately predict the flow of a single soft particle confined inside a microfluidic channel. An outstanding advantage of our numerical developments is that they integrate different types of soft objects in a unique framework, the only difference being in the description of the interfacial mechanics and the time integration schemes. Thanks to the use of Loop elements for the representation of the geometry and unknown fields, the unified formalism established in the weak formulation of this kind of fluid-membrane interaction problem allowed us  to efficiently study confined soft objects spanning from a simple liquid drop to a membrane-enclosed particle with shear or/and bending resistance, such as a capsule, a vesicle, and even a red blood cell.  We have validated the numerical method by comparing the simulation results with highly accurate computations of a liquid drop moving  through tube flow, showing a second-order convergence in both space and time. We have carried out several additional simulations to illustrate the possible applications of the current numerical method, while also demonstrating the accuracy and stability of the algorithm.
Taken together, the code developed here provides a solid base for making a reliable prediction of the dynamical behavior of a confining deformable particle in channel flows, such as the phase diagram, the shape transition, and the lateral migration of 3D vesicles.  These are indeed the subject of our ongoing investigation.

Finally, we point out potential extensions that are closely related to the present work. One of the directions of interest is applying the Loop subdivision to the channel walls only without incorporating the inlet and outlet sections. Such an implementation with a periodic Green's function is highly useful to microfluidic applications and biological flows involving deformable walls. Microfluidic channels are usually fabricated with soft materials which may experience substantial deformations due to the fluid stresses. A deformable wall may be modeled as an elastic shell under the bending-dominated regime in the Kirchhoff–Love equation. A $C^1$ representation of the wall by Loop elements, as developed in the present study,  is then a key to accurately simulating this kind of fluid-structure interaction.
\section*{Declaration of competing interest}

The authors declare that they do not have any conflict of interests.

\section*{Acknowledgments}

We acknowledge financial supports from Labex MEC (grant no. ANR-11-LABX-0092), from A*MIDEX (grant no. ANR-11-IDEX-0001-02), from ANR (grant no. ANR-18-CE06-0008-03), from the LabEx Tec21 (ANR-11-LABX-0030), from the PolyNat Carnot Institute (ANR-11-CARN-007-01) and from CNES. J.M. Lyu was sponsored by the China Scholarship Council (CSC). Centre de Calcul Intensif d’Aix-Marseille is acknowledged for granting access to its high performance computing resources.

%\bibliography{biblio_soft_matter}

\begin{thebibliography}{10}
\expandafter\ifx\csname url\endcsname\relax
  \def\url#1{\texttt{#1}}\fi
\expandafter\ifx\csname urlprefix\endcsname\relax\def\urlprefix{URL }\fi
\expandafter\ifx\csname href\endcsname\relax
  \def\href#1#2{#2} \def\path#1{#1}\fi

\bibitem{Stone_2004}
H.~Stone, A.~Stroock, A.~Ajdari, Engineering flows in small devices:
  {M}icrofluidics toward a lab-on-a-chip, Annu. Rev. Fluid Mech. 36~(1) (2004)
  381--411.
\newblock \href {https://doi.org/10.1146/annurev.fluid.36.050802.122124}
  {\path{doi:10.1146/annurev.fluid.36.050802.122124}}.

\bibitem{Shui_2007}
L.~Shui, J.~C. Eijkel, A.~van~den Berg, Multiphase flow in microfluidic systems
  -- {C}ontrol and applications of droplets and interfaces, Adv. Colloid.
  Interface Sci. 133~(1) (2007) 35--49.
\newblock \href {https://doi.org/10.1016/j.cis.2007.03.001}
  {\path{doi:10.1016/j.cis.2007.03.001}}.

\bibitem{Geislinger_2014}
T.~M. Geislinger, T.~Franke, Hydrodynamic lift of vesicles and red blood cells
  in flow --- from {F}{\aa}hr{\ae}us \& {L}indqvist to microfluidic cell
  sorting, Adv. Colloid. Interface Sci. 208 (2014) 161--176.
\newblock \href {https://doi.org/10.1016/j.cis.2014.03.002}
  {\path{doi:10.1016/j.cis.2014.03.002}}.

\bibitem{WyattShieldsIV_2015}
C.~Wyatt Shields~IV, C.~D. Reyes, G.~P. L{\'o}pez, Microfluidic cell sorting: a
  review of the advances in the separation of cells from debulking to rare cell
  isolation, Lab Chip 15 (2015) 1230--1249.
\newblock \href {https://doi.org/10.1039/C4LC01246A}
  {\path{doi:10.1039/C4LC01246A}}.

\bibitem{SECOMB2013470}
T.~W. Secomb, A.~R. Pries, Blood viscosity in microvessels: Experiment and
  theory, C. R. Phys. 14~(6) (2013) 470--478.
\newblock \href {https://doi.org/10.1016/j.crhy.2013.04.002}
  {\path{doi:10.1016/j.crhy.2013.04.002}}.

\bibitem{Tomaiuolo2009}
G.~Tomaiuolo, M.~Simeone, V.~Martinelli, B.~Rotoli, S.~Guido, Red blood cell
  deformation in microconfined flow, Soft Matter 5 (2009) 3736--3740.
\newblock \href {https://doi.org/10.1039/B904584H}
  {\path{doi:10.1039/B904584H}}.

\bibitem{Liu_2017}
D.~Liu, H.~Zhang, F.~Fontana, J.~T. Hirvonen, H.~A. Santos,
  Microfluidic-assisted fabrication of carriers for controlled drug delivery,
  Lab Chip 17~(11) (2017) 1856--1883.
\newblock \href {https://doi.org/10.1039/C7LC00242D}
  {\path{doi:10.1039/C7LC00242D}}.

\bibitem{Baroud_2010}
C.~N. Baroud, F.~Gallaire, R.~Dangla, Dynamics of microfluidic droplets, Lab
  Chip 10 (2010) 2032--2045.
\newblock \href {https://doi.org/10.1039/C001191F}
  {\path{doi:10.1039/C001191F}}.

\bibitem{Anna_2016}
S.~L. Anna, Droplets and bubbles in microfluidic devices, Annu. Rev. Fluid
  Mech. 48~(1) (2016) 285--309.
\newblock \href {https://doi.org/10.1146/annurev-fluid-122414-034425}
  {\path{doi:10.1146/annurev-fluid-122414-034425}}.

\bibitem{Vlahovska_2013}
P.~M. Vlahovska, D.~Barthes-Biesel, C.~Misbah, Flow dynamics of red blood cells
  and their biomimetic counterparts, C. R. Phys. 14~(6) (2013) 451--458.
\newblock \href {https://doi.org/10.1016/j.crhy.2013.05.001}
  {\path{doi:10.1016/j.crhy.2013.05.001}}.

\bibitem{Barthes_2016}
D.~Barth{\`e}s-Biesel, Motion and deformation of elastic capsules and vesicles
  in flow, Annu. Rev. Fluid Mech. 48~(1) (2016) 25--52.
\newblock \href {https://doi.org/10.1146/annurev-fluid-122414-034345}
  {\path{doi:10.1146/annurev-fluid-122414-034345}}.

\bibitem{Freund_2014}
J.~B. Freund, Numerical simulation of flowing blood cells, Annu. Rev. Fluid
  Mech. 46~(1) (2014) 67--95.
\newblock \href {https://doi.org/10.1146/annurev-fluid-010313-141349}
  {\path{doi:10.1146/annurev-fluid-010313-141349}}.

\bibitem{Trozzo_JCP_2015}
R.~Trozzo, G.~Boedec, M.~Leonetti, M.~Jaeger, Axisymmetric boundary element
  method for vesicles in a capillary, J. Comput. Phys. 289 (2015) 62--82.
\newblock \href {https://doi.org/10.1016/j.jcp.2015.02.022}
  {\path{doi:10.1016/j.jcp.2015.02.022}}.

\bibitem{Chen_2019}
P.~G. Chen, J.~M. Lyu, M.~Jaeger, M.~Leonetti, Shape transition and
  hydrodynamics of vesicles in tube flow, https://arxiv.org/abs/1912.06937
  (2019).

\bibitem{Abreu_2014}
D.~Abreu, M.~Levant, V.~Steinberg, U.~Seifert, Fluid vesicles in flow, Adv.
  Colloid. Interface Sci. 208 (2014) 129--141.
\newblock \href {https://doi.org/10.1016/j.cis.2014.02.004}
  {\path{doi:10.1016/j.cis.2014.02.004}}.

\bibitem{LI2007Augmented}
Z.~Li, K.~Ito, M.-C. Lai, An augmented approach for stokes equations with a
  discontinuous viscosity and singular forces, Comput. Fluids 36 (2007)
  622--635.
\newblock \href {https://doi.org/10.1016/j.compfluid.2006.03.003}
  {\path{doi:10.1016/j.compfluid.2006.03.003}}.

\bibitem{Mendez_2014}
S.~Mendez, E.~Gibaud, F.~Nicoud, An unstructured solver for simulations of
  deformable particles in flows at arbitrary {R}eynolds numbers, J. Comput.
  Phys. 256 (2014) 465--483.
\newblock \href {https://doi.org/10.1016/j.jcp.2013.08.061}
  {\path{doi:10.1016/j.jcp.2013.08.061}}.

\bibitem{YE2015Numerical}
H.~Ye, H.~Huang, X.~yun Lu, Numerical study on dynamic sorting of a compliant
  capsule with a thin shell, Comput. Fluids 114 (2015) 110--120.
\newblock \href {https://doi.org/10.1016/j.compfluid.2015.02.021}
  {\path{doi:10.1016/j.compfluid.2015.02.021}}.

\bibitem{ZHANG2019Lattice}
H.~Zhang, C.~Misbah, Lattice {B}oltzmann simulation of advection-diffusion of
  chemicals and applications to blood flow, Comput. Fluids 187 (2019) 46--59.
\newblock \href {https://doi.org/10.1016/j.compfluid.2019.04.018}
  {\path{doi:10.1016/j.compfluid.2019.04.018}}.

\bibitem{Noguchi2005Shape}
H.~Noguchi, G.~Gompper, Shape transitions of fluid vesicles and red blood cells
  in capillary flows, Proc. Natl. Acad. Sci. USA 102~(40) (2005) 14159--14164.
\newblock \href {https://doi.org/10.1073/pnas.0504243102}
  {\path{doi:10.1073/pnas.0504243102}}.

\bibitem{FedosovCaswell2010b}
D.~A. Fedosov, B.~Caswell, G.~E. Karniadakis, Systematic coarse-graining of
  spectrin-level red blood cell models, Comput. Methods Appl. Mech. Eng. 199
  (2010) 1937--1948.
\newblock \href {https://doi.org/10.1016/j.cma.2010.02.001}
  {\path{doi:10.1016/j.cma.2010.02.001}}.

\bibitem{Lanotte2016red}
L.~Lanotte, J.~Mauer, S.~Mendez, D.~A. Fedosov, J.-M. Fromental, V.~Claveria,
  F.~Nicoud, G.~Gompper, M.~Abkarian, Red cells{\textquoteright} dynamic
  morphologies govern blood shear thinning under microcirculatory flow
  conditions, Proc. Natl. Acad. Sci. USA 113~(47) (2016) 13289--13294.
\newblock \href {https://doi.org/10.1073/pnas.1608074113}
  {\path{doi:10.1073/pnas.1608074113}}.

\bibitem{Pozrikidis_1992}
C.~Pozrikidis, Boundary integral and singularity methods for linearized viscous
  flow, Cambridge {U}niversity {P}ress, 1992.
\newblock \href {https://doi.org/10.1017/CBO9780511624124}
  {\path{doi:10.1017/CBO9780511624124}}.

\bibitem{Lac_JFM2009}
E.~Lac, J.~D. Sherwood, Motion of a drop along the centreline of a capillary in
  a pressure-driven flow, J. Fluid Mech. 640 (2009) 27--54.
\newblock \href {https://doi.org/10.1017/S0022112009991212}
  {\path{doi:10.1017/S0022112009991212}}.

\bibitem{Nagel2015Boundary}
M.~Nagel, F.~Gallaire, Boundary elements method for microfluidic two-phase
  flows in shallow channels, Comput. Fluids 107 (2015) 272--284.
\newblock \href {https://doi.org/10.1016/j.compfluid.2014.10.016}
  {\path{doi:10.1016/j.compfluid.2014.10.016}}.

\bibitem{JoneidI2015Isogeometric}
A.~Joneidi, C.~Verhoosel, P.~Anderson, Isogeometric boundary integral analysis
  of drops and inextensible membranes in isoviscous flow, Comput. Fluids 109
  (2015) 49--66.
\newblock \href {https://doi.org/10.1016/j.compfluid.2014.12.011}
  {\path{doi:10.1016/j.compfluid.2014.12.011}}.

\bibitem{Gounley_JFM_2016}
J.~Gounley, G.~Boedec, M.~Jaeger, M.~Leonetti, Influence of surface viscosity
  on droplets in shear flow, J. Fluid Mech. 791 (2016) 464--494.
\newblock \href {https://doi.org/10.1017/jfm.2016.39}
  {\path{doi:10.1017/jfm.2016.39}}.

\bibitem{Hu_JFM_2012}
X.-Q. Hu, A.-V. Salsac, D.~Barth{\`e}s-Biesel, Flow of a spherical capsule in a
  pore with circular or square cross-section, J. Fluid Mech. 705 (2012)
  176--194.
\newblock \href {https://doi.org/10.1017/jfm.2011.462}
  {\path{doi:10.1017/jfm.2011.462}}.

\bibitem{Boedec_JCP2011}
G.~Boedec, M.~Leonetti, M.~Jaeger, 3{D} vesicle dynamics simulations with a
  linearly triangulated surface, J. Comput. Phys. 230 (2011) 1020--1034.
\newblock \href {https://doi.org/10.1016/j.jcp.2010.10.021}
  {\path{doi:10.1016/j.jcp.2010.10.021}}.

\bibitem{Farutin_JCP_2014}
A.~Farutin, T.~Biben, C.~Misbah, 3{D} numerical simulations of vesicle and
  inextensible capsule dynamics, J. Comput. Phys. 275 (2014) 539--568.
\newblock \href {https://doi.org/10.1016/j.jcp.2014.07.008}
  {\path{doi:10.1016/j.jcp.2014.07.008}}.

\bibitem{Boedec_JCP_2017}
G.~Boedec, M.~Leonetti, M.~Jaeger, Isogeometric {FEM}-{BEM} simulations of
  drop, capsule and vesicle dynamics in {S}tokes flow, J. Comput. Phys. 342
  (2017) 117--138.
\newblock \href {https://doi.org/10.1016/j.jcp.2017.04.024}
  {\path{doi:10.1016/j.jcp.2017.04.024}}.

\bibitem{barakat_shaqfeh_2018b}
J.~M. Barakat, E.~S.~G. Shaqfeh, Stokes flow of vesicles in a circular tube, J.
  Fluid Mech. 851 (2018) 606--635.
\newblock \href {https://doi.org/10.1017/jfm.2018.533}
  {\path{doi:10.1017/jfm.2018.533}}.

\bibitem{Zhao_JCP2010}
H.~Zhao, A.~H.~G. Isfahani, L.~Olson, J.~Freund, A spectral boundary integral
  method for flowing blood cells, J. Comput. Phys. 229~(10) (2010) 3726--3744.
\newblock \href {https://doi.org/DOI: 10.1016/j.jcp.2010.01.024}
  {\path{doi:DOI: 10.1016/j.jcp.2010.01.024}}.

\bibitem{Ramanujan_1998}
S.~Ramanujan, C.~Pozrikidis, Deformation of liquid capsules enclosed by elastic
  membranes in simple shear flow: large deformations and the effect of fluid
  viscosities, J. Fluid Mech. 361 (1998) 117--143.
\newblock \href {https://doi.org/10.1017/S0022112098008714}
  {\path{doi:10.1017/S0022112098008714}}.

\bibitem{Zhong_can_1989}
O.-Y. Zhong-can, W.~Helfrich, Bending energy of vesicle membranes: General
  expressions for the first, second, and third variation of the shape energy
  and applications to spheres and cylinders, Phys. Rev. A 39 (1989) 5280--5288.
\newblock \href {https://doi.org/10.1103/PhysRevA.39.5280}
  {\path{doi:10.1103/PhysRevA.39.5280}}.

\bibitem{Guckenberger_CPC_2016}
A.~Guckenberger, M.~P. Schraml, P.~G. Chen, M.~Leonetti, S.~Gekle, On the
  bending algorithms for soft objects in flows, Comput. Phys. Commun. 207
  (2016) 1--23.
\newblock \href {https://doi.org/10.1016/j.cpc.2016.04.018}
  {\path{doi:10.1016/j.cpc.2016.04.018}}.

\bibitem{Cirak_2000}
F.~Cirak, M.~Ortiz, P.~Schr{\"o}der, Subdivision surfaces: a new paradigm for
  thin-shell finite-element analysis, Int. J. Numer. Meth. Eng. 47~(12) (2000)
  2039--2072.
\newblock \href
  {https://doi.org/10.1002/(SICI)1097-0207(20000430)47:12<2039::AID-NME872>3.0.CO;2-1}
  {\path{doi:10.1002/(SICI)1097-0207(20000430)47:12<2039::AID-NME872>3.0.CO;2-1}}.

\bibitem{Hughes2005Isogeometric}
T.~Hughes, J.~Cottrell, Y.~Bazilevs, Isogeometric analysis: {CAD}, finite
  elements, {NURBS}, exact geometry and mesh refinement, Comput. Methods Appl.
  Mech. Eng. 194~(39) (2005) 4135--4195.
\newblock \href {https://doi.org/10.1016/j.cma.2004.10.008}
  {\path{doi:10.1016/j.cma.2004.10.008}}.

\bibitem{Loop_1987}
C.~Loop, Smooth subdivision surfaces based on triangles, Master's thesis,
  Department of Mathematics, University of Utah (1987).

\bibitem{Maestre_CMAME_2017}
J.~Maestre, J.~Pallares, I.~Cuesta, M.~A. Scott, A 3{D} isogeometric {BE}--{FE}
  analysis with dynamic remeshing for the simulation of a deformable particle
  in shear flows, Comput. Methods Appl. Mech. Eng. 326 (2017) 70--101.
\newblock \href {https://doi.org/10.1016/j.cma.2017.08.003}
  {\path{doi:10.1016/j.cma.2017.08.003}}.

\bibitem{Bartezzaghi_CMAME_2019}
A.~Bartezzaghi, L.~Ded{\`e}, A.~Quarteroni, Biomembrane modeling with
  isogeometric analysis, Comput. Methods Appl. Mech. Eng. 347 (2019) 103--119.
\newblock \href {https://doi.org/10.1016/j.cma.2018.12.025}
  {\path{doi:10.1016/j.cma.2018.12.025}}.

\bibitem{barakat_shaqfeh_2019}
J.~M. Barakat, S.~M. Ahmmed, S.~A. Vanapalli, E.~S.~G. Shaqfeh, Pressure-driven
  flow of a vesicle through a square microchannel, J. Fluid Mech. 861 (2019)
  447--483.
\newblock \href {https://doi.org/10.1017/jfm.2018.887}
  {\path{doi:10.1017/jfm.2018.887}}.

\bibitem{Helfrich_1973}
W.~Helfrich, Elastic properties of lipid bilayers: theory and possible
  experiments, Z. Naturforsch. 28c (1973) 693--703.

\bibitem{Lyu_2018}
J.~Lyu, P.~G. Chen, G.~Boedec, M.~Leonetti, M.~Jaeger, Hybrid
  continuum--coarse-grained modeling of erythrocytes, C. R. Mec. 346 (2018)
  439--448.
\newblock \href {https://doi.org/10.1016/j.crme.2018.04.015}
  {\path{doi:10.1016/j.crme.2018.04.015}}.

\bibitem{Stam_1998}
J.~Stam, Evaluation of loop subdivision surfaces, SIGGRAPH 99 Course Notes
  (2001).

\bibitem{Cirak_2011}
F.~Cirak, Q.~Long, Subdivision shells with exact boundary control and
  non-manifold geometry, Int. J. Numer. Meth. Eng. 88~(9) (2011) 897--923.
\newblock \href {https://doi.org/10.1002/nme.3206}
  {\path{doi:10.1002/nme.3206}}.

\bibitem{Walter_2010}
J.~Walter, A.-V. Salsac, D.~Barth{\`e}s-Biesel, P.~Le~Tallec, Coupling of
  finite element and boundary integral methods for a capsule in a stokes flow,
  Int. J. Numer. Meth. Eng. 83~(7) (2010) 829--850.
\newblock \href {https://doi.org/10.1002/nme.2859}
  {\path{doi:10.1002/nme.2859}}.

\bibitem{Pozrikidis_2005}
C.~Pozrikidis, Numerical simulation of cell motion in tube flow, Ann. Biomed
  Eng. 33~(2) (2005) 165--178.
\newblock \href {https://doi.org/10.1007/s10439-005-8975-6}
  {\path{doi:10.1007/s10439-005-8975-6}}.

\bibitem{Fehlberg_1969}
E.~Fehlberg, Low-order classical {R}unge-{K}utta formulas with stepsize control
  and their application to some heat transfer problems, Tech. rep., NASA
  (1969).

\bibitem{Knoll_2004}
D.~A. Knoll, D.~E. Keyes, Jacobian-free {N}ewton--{K}rylov methods: a survey of
  approaches and applications, J. Comput. Phys. 193~(2) (2004) 357--397.
\newblock \href {https://doi.org/10.1016/j.jcp.2003.08.010}
  {\path{doi:10.1016/j.jcp.2003.08.010}}.

\bibitem{Yih_1979}
C.-S. Yih, Fluid Mechanics, West River Press, 1979.

\bibitem{Kuriakose_2011}
S.~Kuriakose, P.~Dimitrakopoulos, Motion of an elastic capsule in a square
  microfluidic channel, Phys. Rev. E 84 (2011) 011906.
\newblock \href {https://doi.org/10.1103/PhysRevE.84.011906}
  {\path{doi:10.1103/PhysRevE.84.011906}}.

\bibitem{Liron_1978}
N.~Liron, R.~Shahar, Stokes flow due to a {S}tokeslet in a pipe, J. Fluid Mech.
  86 (1978) 727--744.
\newblock \href {https://doi.org/10.1017/S0022112078001366}
  {\path{doi:10.1017/S0022112078001366}}.

\bibitem{Brenner_1970}
H.~Brenner, Pressure drop due to the motion of neutrally buoyant particles in
  duct flows, J. Fluid Mech. 43 (1970) 641--660.
\newblock \href {https://doi.org/10.1017/S0022112070002641}
  {\path{doi:10.1017/S0022112070002641}}.

\bibitem{Hetsroni_JFM_1970}
G.~Hetsroni, S.~Haber, Wacholder, The flow fields in and around a droplet
  moving axially within a tube, J. Fluid Mech. 41~(4) (1970) 689--705.
\newblock \href {https://doi.org/10.1017/S0022112070000848}
  {\path{doi:10.1017/S0022112070000848}}.

\bibitem{Hu_2013}
X.-Q. Hu, B.~S\'ev\'eni\'e, A.-V. Salsac, E.~Leclerc, D.~Barth\`es-Biesel,
  Characterizing the membrane properties of capsules flowing in a
  square-section microfluidic channel: Effects of the membrane constitutive
  law, Phys. Rev. E 87 (2013) 063008.
\newblock \href {https://doi.org/10.1103/PhysRevE.87.063008}
  {\path{doi:10.1103/PhysRevE.87.063008}}.

\bibitem{Farutin_2014}
A.~Farutin, C.~Misbah, Symmetry breaking and cross-streamline migration of
  three-dimensional vesicles in an axial {P}oiseuille flow, Phys. Rev. E 89
  (2014) 042709.
\newblock \href {https://doi.org/10.1103/PhysRevE.89.042709}
  {\path{doi:10.1103/PhysRevE.89.042709}}.

\bibitem{agarwal2020stable}
D.~Agarwal, G.~Biros, Stable shapes of three-dimensional vesicles in unconfined
  and confined {P}oiseuille flow, Phys. Rev. Fluids 5~(1) (2020) 013603.
\newblock \href {https://doi.org/10.1103/PhysRevFluids.5.013603}
  {\path{doi:10.1103/PhysRevFluids.5.013603}}.

\bibitem{Evans_1980}
E.~A. Evans, R.~Skalak, Mechanics and thermodynamics of biomembranes, CRC
  Press, 1980.

\end{thebibliography}

\end{document}